\documentclass[sigconf]{acmart}

\copyrightyear{2024}
\acmYear{2024}
\setcopyright{acmlicensed}\acmConference[CCS '24]{Proceedings of the 2024 ACM SIGSAC Conference on Computer and Communications Security}{October 14--18, 2024}{Salt Lake City, UT, USA}
\acmBooktitle{Proceedings of the 2024 ACM SIGSAC Conference on Computer and Communications Security (CCS '24), October 14--18, 2024, Salt Lake City, UT, USA}
\acmDOI{10.1145/3658644.3690290}
\acmISBN{979-8-4007-0636-3/24/10}

\usepackage[]{wspr}

\usepackage{amsmath}
\usepackage{amsthm}
\usepackage{booktabs}
\usepackage{amsfonts}
\usepackage{caption}
\usepackage{diagbox}
\usepackage[group-separator={,}]{siunitx}
\usepackage{subcaption}
\usepackage{algorithmicx}
\usepackage{adjustbox}
\usepackage{graphics}
\usepackage{graphicx}
\usepackage{tikz}
\usepackage{url}
\usepackage{pifont}

\usepackage{graphicx}
\usepackage{url}

\usepackage{algorithm}
\usepackage{algpseudocode}

\newif\ifisFullVersion

\isFullVersiontrue

\ifisFullVersion
\usepackage{paralist}
\usepackage[normalem]{ulem}
\useunder{\uline}{\ul}{}
\fi

\usepackage[
	lambda,
	operators,
	advantage,
	sets,
	adversary, 
	landau,
	probability,
	notions,
	logic,
	ff,
	mm,
	primitives,
	events,
	complexity,
	asymptotics,
	keys]{cryptocode}

\usepackage{varioref}
\usepackage{hyperref}
\hypersetup{
  linkcolor  = red,
  citecolor  = green,
  urlcolor   = blue,
  colorlinks = true,
  allcolors=blue
}
\usepackage[nameinlink,capitalize]{cleveref}
\usepackage{enumitem}

\usepackage[most]{tcolorbox}
\usepackage[section]{placeins}

\newcommand{\projectname}{J\"ager\xspace}
\newcommand{\sysname}{\projectname}
\newcommand{\FFF}{\mathcal{F}}
\newcommand{\fprivtb}{\FFF_{{\sf TraceBack}}}

\newcommand{\remove}[1]{}

\newcommand{\WE}{\mathsf{WE}}
\mathchardef\mhyphen="2D

\newcommand{\ct}{\mathsf{ct}}

\newcommand{\gpk}{\mathsf{gpk}}

\newcommand{\gsk}{\mathsf{gsk}}

\newcommand{\DDD}{\mathcal{D}}
\newcommand{\LLL}{\mathtt{L}}
\newcommand{\GGG}{\mathcal{G}}
\newcommand{\ggs}{\GGG_\mathsf{gsign}}
\newcommand{\enroll}{\mathtt{gENROLL}}
\newcommand{\vfysign}{\mathtt{gVERIFY}}
\newcommand{\info}{\mathsf{info}}

\newcommand{\validate}{\mathsf{Validate}}
\newcommand{\trace}{\mathsf{trace}}
\newcommand{\tmax}{t_\mathsf{max}}
\newcommand{\cdrkey}{\mathsf{key}}

\newcommand{\ctime}{\mathsf{record}\text{-}\mathsf{index}}

\newcommand{\malcdr}{\mathcal{C}_\mathsf{malicious}}
\newcommand{\CCC}{\mathcal{C}}

\newcommand{\km}{\mathsf{LM}}
\newcommand{\itg}{\TP}

\newcommand{\cregister}{\mathtt{REGISTER}}
\newcommand{\PPP}{\mathcal{P}}
\newtheorem{theorem}{Theorem}

\newcommand{\MMM}{\mathcal{M}}

\newcommand{\rlctr}{\mathsf{rl}\text{-}\mathsf{ctr}}
\newcommand{\idx}{\mathsf{idx}}
\newcommand{\rlimit}{\mathsf{T}}

\newcommand{\currtime}{\mathsf{curr}\text{-}\mathsf{time}}

\newcommand{\provider}{{P}}
\newcommand{\carrier}{provider\xspace}
\newcommand{\carriers}{providers\xspace}
\newcommand{\Carriers}{Providers\xspace}
\newcommand{\StirSha}{S/S\xspace}
\newcommand{\src}{src}
\newcommand{\dst}{dst}
\newcommand{\ts}{ts}

\newcommand{\oprf}{\mathsf{OPRF}}
\newcommand{\gsetup}{\mathbf{GSetup}}
\newcommand{\gkgen}{\mathbf{GKGen}}
\newcommand{\gjoin}{\mathbf{Join}}
\newcommand{\updgrp}{\mathbf{UpdateGroup}}
\newcommand{\gsign}{\mathbf{Sign}}
\newcommand{\gverify}{\mathbf{Verify}}
\newcommand{\gopen}{\mathbf{Open}}
\newcommand{\foprf}{\mathcal{F}_{\mathsf{OPRF}}}
\newcommand{\ep}{\mathsf{ep}}
\newcommand{\copen}{\mathtt{OPEN}}
\newcommand{\delete}{\mathtt{del}}
\newcommand{\add}{\mathtt{add}}
\newcommand{\allowtrace}{\mathtt{ALLOW}\text{-}\mathtt{TRACE}}
\newcommand{\malupdate}{\mathtt{MAL}\text{-}\mathtt{UPDATE}}

\newcommand{\cdr}{\mathsf{hop}}
\newcommand{\callpp}{\mathsf{call}\text{-}\mathsf{details}}

\newcommand{\clabel}{\mathsf{call}\text{-}\mathsf{label}}
\newcommand{\ttoep}{\mathsf{tsToEpoch}}

\newcommand{\retrieve}{\mathtt{RETRIEVE}}
\newcommand{\ta}{\TA}
\renewcommand{\LLL}{\mathcal{L}}
\newcommand{\retreq}{\mathtt{RETRIEVE}\text{-}\mathtt{REQ}}

\newcommand{\conflicts}{\mathsf{conflicts}}
\newcommand{\fro}{\mathcal{F}_{\mathsf{RO}}}

\newcommand{\QQQ}{\mathcal{Q}}
\newcommand{\ccall}{\mathtt{CONTRIBUTE}}
\newcommand{\ctrace}{\mathtt{TRACE}}

\newcommand{\ok}{\mathtt{OK}}

\newcommand{\prfinit}{\mathtt{INIT}}
\newcommand{\prfeval}{\mathtt{EVAL}}
\newcommand{\prfproceed}{\mathtt{PROCEED}}

\newcommand{\hyb}{\mathbf{Hybrid}}

\newcommand{\prffail}{\mathsf{PRFFail}}
\newcommand{\rofail}{\mathsf{ROFail}}
\newcommand{\sigfail}{\mathsf{SigFail}}
\newcommand{\groupsigfail}{\mathsf{GroupSigFail}}
\newcommand{\framefail}{\mathsf{FrameFail}}
\newcommand{\veroprffail}{\mathsf{VerifyOPRFFail}}

\newcommand{\RS}{RS\xspace}
\newcommand{\GM}{GM\xspace}

\newcommand{\TA}{TA\xspace}

\newcommand{\TP}{{\RS}}
\newcommand{\protosetup}{Setup\xspace}
\newcommand{\protocontribute}{Contribution\xspace}
\newcommand{\prototrace}{Trace\xspace}
\newcommand{\protoopen}{Provider Accountability\xspace}

\newcommand{\ale}[1]{\notetext{\textcolor{magenta}{Ale: #1}}}

\newcommand{\myparagraph}[1]{\vspace{4pt}\noindent\textbf{#1}:}
\newcommand{\negvspace}{\vspace{-4pt}}
\newcommand{\smallspace}{\vspace{2pt}}
 
\title{ \projectname: Automated Telephone Call Traceback}

\newcommand{\ncsuemail}[1]{\email{#1@ncsu.edu}}

\newcommand{\ncsuaffil}{\affiliation{\institution{North Carolina State University}
    \city{Raleigh}
    \country{USA}
  }
}

\author{David Adei}
\orcid{0009-0007-9857-6671}
\ncsuaffil
\ncsuemail{dahmed}

\author{Varun Madathil}
\orcid{0009-0001-4390-3096}
\ncsuaffil
\ncsuemail{vrmadath}

\author{Sathvik Prasad}
\orcid{0009-0002-1862-9483}
\ncsuaffil
\ncsuemail{snprasad}

\author{Bradley Reaves}
\orcid{0000-0001-7902-1821}
\ncsuaffil
\ncsuemail{bgreaves}

\author{Alessandra Scafuro}
\orcid{0000-0003-1797-9457}
\ncsuaffil
\ncsuemail{ascafur} \renewcommand{\shortauthors}{David Adei, Varun Madathil, Sathvik Prasad, Bradley Reaves, \& Alessandra Scafuro}
\begin{document}

\begin{abstract}
Unsolicited telephone calls that facilitate fraud or unlawful telemarketing continue to overwhelm network users and the regulators who prosecute them.
The first step in prosecuting phone abuse is traceback --- identifying the call originator. This fundamental investigative task currently requires hours of manual effort per call.
In this paper, we introduce \sysname, a distributed secure call traceback system.
\sysname can trace a call in a few seconds, even with partial deployment, while cryptographically preserving the privacy of call parties, carrier trade secrets like peers and call volume, and limiting the threat of bulk analysis.
We establish definitions and requirements of secure traceback, then develop a suite of protocols that meet these requirements using witness encryption, oblivious pseudorandom functions, and group signatures. We prove these protocols secure in the universal composibility framework.
We then demonstrate that \sysname has low compute and bandwidth costs per call, and these costs scale linearly with call volume.
\sysname provides an efficient, secure, privacy-preserving system to revolutionize telephone abuse investigation with minimal costs to operators.
\end{abstract}
 



\maketitle

\section{Introduction}
\label{sec:introduction}

Telephone networks are inundated with unsolicited ``robocalls'' for telemarketing
or outright fraud. 
While individuals are bothered by the seemingly constant ringing of the phone,
government agencies, enterprises, and non-profits now have greater
difficulties reaching stakeholders for legitimate, desirable purposes.
Public outcry has motivated policy makers, public
officials, and phone providers to all take actions~\cite{fcc_traced_act,itg} meant to 
address the problem.

Prior to the recent spate of incessant robocalls, the Federal Trade Commission (FTC) established the Telemarketing Sales Rule~\cite{tsr_ftc}
to set the bounds on what forms of automated calls are considered permissible.
A general summary is that callers
must have affirmative, opt-in consent from the called party to dial them for 
commercial purposes. The FTC also maintains a ``do-not-call'' list that 
should prevent unsolicited calls to registered individuals. These measures are
legal, not technical, so violations are pursued through regulatory
action. These measures have clearly not been effective.

In late 2019, a sudden outbreak of bipartisanship struck the United States
Congress, who passed the TRACED Act to combat illegal calling. Among other
measures, the law further expanded penalties for illegal calling and empowered
regulators to make substantial changes to network policy to reduce robocalls.
These changes to date have included requiring providers to register and submit
Robocall Mitigation Plans to the FCC, the mandatory blocking of calls claiming
to originate from invalid numbers, encouraging the labeling of suspect calls
by providers, setting deadlines on participation in robocall investigations, 
and mandating that all providers implement a call authentication mechanism
known as STIR/SHAKEN (\StirSha). \StirSha requires originating providers to sign
outbound call requests to indicate their actual source, similar to DKIM, expecting that it would allow regulators to identify the source of the call,
prevent caller ID spoofing, and give robocallers ``no place to hide.''

In practice, all of these efforts have failed to significantly change the
state of affairs. Call labelling is unreliable, robocallers have moved to 
using legitimate numbers for a very short period, and the majority of calls in
the network arrive without a signature~\cite{transnexusstiroctstats} because pre-VoIP networks cannot be
modified to support \StirSha.

Robocallers continue to operate for a simple reason: it is profitable and low
risk. While regulators and law enforcement have been successful in winning
judgements against accused robocall operators, with fines into many millions
of dollars, the defendants often return to their schemes under assumed
identities or are replaced by other ``entrepreneuers'' using similar
techniques. Because the robocalling problem is so vast, it is reasonable to
assume that they will not face penalties given how painstaking it is to bring
a case against a single robocaller and how small the relevant agency staffs are.

One of the biggest hurdles is identifying the source of a call. The
telephone network is a network-of-networks, like the Internet, and a given
call often passes through many networks before reaching its
destination. Routes change rapidly and unpredictably, and
providers routing the call only know the previous ``hop'' and the next
``hop.'' Fortunately, providers keep meticulous records on calls they
route for billing or paying peers. To identify the
source of a call, an investigator must start at the destination with
knowledge of the call time, call destination, and claimed call source, and go 
hop-by-hop until they reach the originator. This process is termed
\textbf{traceback}.

Prior to 2019, a traceback, could take multiple subpoenas and months to complete \emph{for a
single call}~\cite{bercu23}. The 2019 TRACED Act mandated the creation of a
clearinghouse to handle tracebacks, and it also mandated timely
responses to traceback requests by providers. The result was the creation of
the Industry Traceback Group (ITG)~\cite{itg}, and the FCC reported to Congress that
currently a traceback can be completed in under 24 hours with the help of ITG.

Unfortunately, tracebacks are
still largely manual, and the ITG has a skeleton staff of a few employees.
They are (rightfully) proud that they manage to complete around 300
tracebacks per month, though this is far from sufficient
to deal with millions of robocalls each month. It will certainly become
a bottleneck if we want law enforcement to target perpetrators of high-touch
fraud schemes like digital kidnapping, where a fraudster provides a convincing
story and fabricated voice of the loved one of a target to extort a ransom.

Automation is needed to scale traceback to a significant fraction
of the current abuse, but simple approaches will be unacceptable to some
portion of current stakeholders. If each carrier were required to implement an
API for traceback, it runs the risk of a malicious carrier fabricating
incorrect responses. Note that some smaller carriers are known to skirt or
outright violate laws for profit, including facilitating illegal robocalls. If
each provider were required to submit their call routing records to a central
source, subscribers would justifiably worry that their social networks and
telephone activity could be leaked. Providers would balk at revealing 
peering arrangements and call volumes, and the central database would be a
magnet for curious intelligence agencies and law enforcement dragnets.

In this paper, we present \sysname\footnote{\sysname, pronounced ``YAYger,'' is German for ``hunter.'' It can also refer to J\"agermeister, a popular liqueur, making it an appropriate name for a protocol to supplement \StirSha.}, a distributed system and protocol suite to
provide rapid, automated traceback for phone calls. To trace an illegal call, an investigator will obtain the caller's and recipient's telephone numbers along with precise timing details about the call. This information is typically sourced from public complaints, industry honeypots, the ITG's data collection, consumer voicemails, or other commercial channels. \sysname enables the investigator with the call details to identify its originating network. The originating network can then be held liable or identify the customer responsible for the illegal call.

The key insight is to
allow traceback over encrypted call records, with the caveat that only a party
with detailed knowledge of the call can identify or decrypt the routing records for a given call. Our solution
requires no modifications to the existing telephone network. Instead, we
assume access to the billing systems that already maintain the records we
need. We do not require interaction between providers at any point. The
compute, storage, and bandwidth costs for providers are modest and scale
linearly with call volume, so small carriers need few resources. 
We provide
for cryptographic mechanisms to ensure that authorized traceback users can be
appropriately rate-limited to prevent bulk abuse. 
Our system is robust against
a single provider on a call who fabricates or does not submit a routing record
for a call. Encrypted call records do not reveal the provider who submitted
them, but a provider who submits an invalid or incorrect record can be
identified. 
Because the purpose of traceback is to identify the source of a call, we do not actually need every routing record for a call to be present in \sysname. Ideally, at least the first, originating record, will be present, but if it isn't, any other record will still improve traceback performance.
Our scheme will provide benefits even if some providers do not
participate, so it can be deployed incrementally.

We make the following contributions:

\begin{itemize}[itemsep=0pt, topsep=2pt]

\item We specify and define the key properties and requirements for secure
   telephone call traceback.

\item We design protocols that meet the requirements for secure
traceback, and implement them in a prototype distributed system dubbed
\sysname. 

\item We provide formal guaranties by proving the security of these protocols in the Universal Composability
(UC) framework.

\item We demonstrate that \sysname has low compute and bandwidth costs per call, and these costs scale linearly with call volume. In the process, we
also develop a performant witness encryption library in C++. Code for that library and our full \sysname implementation is available~\cite{jagerimpgit2024, jagerimpzen2024}.

\end{itemize}
Robocall enforcement is ultimately a legal problem, albeit with technical challenges. \sysname fills a technical need for effective investigative tools, though \sysname, cannot independently determine whether a call was illegal.
Nevertheless, because the telephone ecosystem is heavily regulated, honest participation can be incentivized through the risk of civil or criminal prosecution.

 \section{Background}
\label{sec:background}

This section discusses background information on the state of the telephone network abuse and prosecuting violators. In doing so, we described challenges with locating abuse actors.

\subsection{Telephone Network Abuse}
\label{subsec:telephonenetworkabuse}

Phone network abuse is a global problem, with the United States being one of the most severely 
affected countries.
One of the most common forms of abuse is pre-recorded automated bulk phone calls, or robocalls.   
Many robocalls violate US law, including sales calls made without affirmative opt-in.
Fraudulent calls often impersonate government officials and steal millions of dollars from victims.

Federal statues prescribe eyewatering financial penalties for each and every illegal call,
but penalties require enforcement to deter abuse. Currently, 
phone abusers avoid prosecution by using spoofed or short-term telephone numbers, regularly
changing service providers, and altogether vastly exceeding available enforcement resources.

\myparagraph{Routing Phone Calls through the Network}\label{sec:route-calls}
Determining the source of a single call currently requires significant
investigative effort, and part of the reason is that the phone network
is a network-of-networks with no global vantage point and no single end-to-end
authentication of identity.

A given telephone provider will connect with one or more other providers to 
send and receive call traffic. 
When a subscriber places a call, her provider ``originates'' the call and then 
uses a signalling protocol to communicate with its peer networks to find a route
to the called party's network. When the call is finally set-up, over potentially many intermediate \carriers,  the call is considered
``terminated'' and the call audio will begin.

Call routes are selected considering carrier\footnote{In this work, we will use carrier and provider interchangeably.} charges, network maintenance, and agreements with other carriers, and these factors change moment-to-moment.
Each provider that carries the call will bill the provider who sent it, and Call Detail Records (CDRs) are kept to support this. However, only the originating provider knows any details about the call originator beyond the phone number the subscriber claimed when they set up the call.\footnote{Most businesses expect to be able to specify the ``from'' field shown for caller ID. A common case is to allow a desk line to appear to be coming from the corporate switchboard number, but this feature is abused by illegal callers.} No provider ever knows more about the call than the previous and next provider in the route. To identify a party responsible for an illegal call, an investigator must first find the originating provider, which is unknown to the terminating provider who delivered the call to its recipient.

\myparagraph{Locating Abuse Actors}\label{subsec:locatingabuseactors}
Authorities must find and prosecute perpetrators responsible for generating illegal calls.
In the United States, the TRACED Act of 2019~\cite{fcc_traced_act} requires the FCC to mandate the \StirSha caller authentication framework. 
Although in principle S/S can be used to traceback illegal robocalls from their destination to their origin, there are several limitations.
Industry reports from October 2023 estimate~\cite{transnexusstiroctstats} more than half of all voice traffic in the US is still not signed using S/S, making it impossible to track the origin of such calls using S/S information alone.
Industry insiders attribute this large portion of unauthenticated voice traffic to legacy infrastructure that does not support S/S. 
Furthermore, phone calls originating outside the US often do not contain S/S information since the framework is not mandated in other countries.
Therefore, regulators, enforcement agencies, and other entities rely exclusively on manual traceback processes to identify the source of illegal robocalls~\cite{fcc_investigation_1}.

\myparagraph{Manual Traceback Process}\label{subsec:manualtracebackprocess}
As the TRACED Act requires, the FCC has designated the ITG~\cite{itg} to serve as the central entity to coordinate the traceback process.
The ITG manages the labor-intensive and time-consuming tasks of identifying the source of suspected illegal robocalls.
The ITG constantly monitors active robocall campaigns using data from honeypots~\cite{usenix_whoscalling}, consumer reports, and other sources.
After assessing the legality of the robocall, the ITG initiates a traceback request and manually coordinates across numerous carriers to pinpoint the source of suspected illegal robocalls.
The traceback process involves tracing the call path from the terminating carrier through transit carriers and ultimately to the originating carrier to identify the source of the call. 

Traceback has proven to be a crucial tool for regulators and enforcement agencies to combat illegal robocalls. It has been used in almost every enforcement action filed by regulators against robocalling operations.
However, successfully completing a traceback often takes several hours or days, with substantial effort from the ITG and the participating carriers.
The manual and time-consuming nature of the traceback process significantly limits its effectiveness.
Although the volume of illegal robocalls targeting US subscribers is estimated to be in the hundreds of millions per year, less than 3,000 tracebacks were completed over eleven months in 2022~\cite{traceback_stats}.
By developing an automated, secure, and scalable traceback system, we can swiftly uncover the source of such calls, deter bad actors, and empower stakeholders to protect phone users from illegal robocalls.

\subsection{Cryptographic Primitives}
\label{subsec:protocol-prelims}
This section introduces the cryptographic primitives that serve as the building blocks for our protocol.

\myparagraph{Witness Encryption Based on Signatures} A Witness Encryption scheme based on Signatures (WES) was recently proposed in~\cite{wes} and~\cite{mcfly}. These are encryption schemes where the encryption key is a tuple of a signature verification key (denoted $\vk$) and a string chosen by the encryptor (denoted $\ell$).  The decryption key is a valid signature (denoted $\sigma$) on the string, such that the signature can be verified by the verification key. 
More specifically, a witness encryption based on signatures has two algorithms - $\WE.\enc, \WE.\dec$, where $\WE.\enc((\vk, \ell), m) \rightarrow \ct$, and $\WE.\dec(\sigma, \ct) \rightarrow m$. $m$ denotes the plaintext, $(\vk, \ell)$ corresponds to the encryption key, and $\sigma = \sign(\sk, \ell)$ corresponds to the decryption key if $\sign.\verify(\vk, \ell, \sigma) = 1$. Here $\sign(\cdot\cdot\cdot)$ and $\sign.\verify(\cdot\cdot\cdot)$ are the sign and verify procedures for the signature scheme.  
We acknowledge that using signature verification keys to encrypt messages and using signatures to decrypt ciphertexts is not intuitive, and the notation can be confusing. Observe that we denote witness encryption and decryption keys as \emph{tuples} containing signing and verification keys, while the signature keys are written as single variables. 
~\cite{wes} and~\cite{mcfly} show that it is possible to construct such WES schemes efficiently based on BLS signatures~\cite{boneh2019bls}.

\myparagraph{Group Signatures} Group signatures~\cite{ateniese2005practical} are a cryptographic primitive that allows group members to anonymously sign messages on behalf of the group. 
A designated authority, the group manager, generates a common public key $\gpk$ and issues a unique group member signing key $\gsk_i$ for each group member $i$. Any signature signed by any $\gsk_j$ in the group will verify with $\gpk$. The group manager can also deanonymize signatures and identify the signer. Group signatures allow for anonymity while maintaining accountability. 

\myparagraph{Oblivious PRF} A pseudorandom function (PRF)  $F_k$ is a keyed function whose outputs look random to anyone without the secret key $k$. An oblivious pseudorandom function (OPRF)~\cite{casacuberta2022sok} is a two-party protocol where a server holds a secret key $k$ for the PRF, and a client holds a secret input $x$ to be evaluated. At the end of the protocol, the client learns $F_k(x)$ while the server learns nothing.

\section{Problem Statement}
\label{sec:problem}
We begin this section by identifying the major stakeholders and adversaries impacting \sysname. We also specify the functional and security requirements which we aim to achieve.

\subsection{Stakeholders and Adversaries}

The ecosystem involves various stakeholders with distinct roles and interests, including service providers, the ITG, subscribers, and law enforcement agencies. Each group's goals and actions follow.

\myparagraph{Providers} They route phone calls through the telephone network. By law~\cite{fcc_cdr_mandate}, they are required to maintain the CDRs of each call and actively participate in the traceback process. They prioritize efficient record insertion and complete and correct traceback responses. Providers also desire the confidentiality of their customers, peering partners, and traffic volumes.

\myparagraph{Subscribers} Subscribers initiate and receive phone calls. They also report fraudulent calls to authorities or their service provider. Subscribers seek to minimize receiving illegal robocalls and expect confidentiality for their call records.

\myparagraph{Industry Traceback Group} The ITG oversees the tracing of illegal calls to their source by working with providers.
They prioritize swift and accurate responses to traceback requests.

\myparagraph{Regulatory and Law Enforcement Agencies (LEAs)} LEAs work in conjunction with industry stakeholders to maintain secure and lawful communication networks. Their responsibilities include investigating suspected illegal calls, enforcing compliance, public education, and policy development. LEAs often submit traceback requests to the ITG,  emphasizing the need for a timely response.

\myparagraph{Adversaries}
Adversaries may seek partial or full call records of one, many, or all subscribers. They may also seek  privileged information about providers or the network structure. They may also aim to violate the integrity or availability of \sysname to prevent detection or investigation of illegal calls.
Adversaries can include outside parties like private investigators\cite{heuser2017phonion}, identity thieves, or even foreign intelligence agencies\cite{nsa_spying}. Insiders, including subscribers, providers, regulators and LEAs, and operators of \sysname entities may also
behave dishonestly at any point. We design \sysname such that no single compromised entity alone can violate its security properties, and in many cases \sysname is resilient against collusion by more than one malicious entity. 

\subsection{Requirements}
\label{sec:requirements}
The main objective of \sysname is to enable secure and efficient traceback given a valid request containing source and destination telephone numbers along with the call timestamp.

\myparagraph{Functional Requirements} To achieve this objective, the system is designed with the following key requirements: 

\begin{enumerate}[itemsep=0pt, topsep=2pt]
    \item \noindent\underline{\textit{Resilience}}: A valid traceback request returns all available records for a call.
    
    \item \noindent\underline{\textit{Precision}}: A valid traceback request only returns relevant records. Malicious or incorrect records are still ``relevant'' if they match the traceback request. 

    \item\noindent\underline{\textit{Scalability}}: \sysname must handle effectively arbitrary call volumes. For all entities, cost should scale linearly in the number of calls and/or participants (as appropriate). 

    \item \noindent\underline{\textit{Efficiency}}: All operations should perform comparably to similar non-secure approaches. The financial costs should be a minor fraction of the total network revenue. 

    \item \noindent\underline{\textit{Information Gain}}: Every traceback request should provide information to an investigator. A traceback request will result in one or more of the following:
    \begin{enumerate}[itemsep=0pt, topsep=0pt]
        \item Identify the originating provider for the call.
        \item Reveal at least one claimed non-originating provider and shorten manual traceback.
        \item Provide direct evidence that one or more providers act in bad faith (e.g., submitting false or contradictory records or no records for a call).
    \end{enumerate}
    In settings where \sysname is mandatory, all of these properties are obvious. Either one obtains a complete and consistent traceback, or at least one provider is violating the mandate. In partial deployment, these properties will still hold if at least one on-path provider participates.
\end{enumerate}

\myparagraph{Security Requirements} 
The guiding principle of secure traceback should be that no entity gains information about subscribers or providers in the absence of an authorized traceback request, even in the presence of a compromised \sysname entity.
Additionally, no entity can provide false information without risk of detection and accountability. More formally, this mandates the following principles:

\begin{enumerate}[itemsep=0pt, topsep=2pt]
    \item\noindent\underline{\textit{Trace authorization}}: An entity can only trace a call
    they have definite knowledge of, and they must also have
    explicit authorization from a third party.

    \item\noindent\underline{\textit{Call confidentiality}}: No entity should determine source, destination, time, or route details about a call they do not already have without authorization for a traceback. 

    \item\noindent\underline{\textit{Trade secret protection}}: No party should learn aggregate information about a provider's call volumes and peering relations except those revealed by an authorized, valid traceback or an authorized accountability request. 

    \item\noindent\underline{\textit{Record integrity}}: Only authorized parties may contribute records.

    \item\noindent\underline{\textit{Record accountability}}: It must be possible to identify the contributor of a traceback record. 
\end{enumerate}

 \section{Our Approach}
\label{sec:approach}

In the previous section, we specified requirements for secure traceback.
To show how \sysname satisfies those requirements, in this section
we will describe a functional but insecure strawman solution and iteratively improve it until it meets all of the security requirements.

\subsection{\sysname Overview}
\label{sec:strawman}
\negvspace\myparagraph{An Insecure Strawman Approach}
To enable traceback, we first introduce a central Record Store(\RS) that collects and stores Call Detail Records (CDRs) from \carriers{} in a database $\DDD$. Any \carrier{} $P_i$ in a call path, for instance, $P_1 \rightarrow P_2 \rightarrow P_3 \rightarrow P_4$, already keeps a CDR for each call they originate, transmit, or terminate. We model a CDR  as a tuple  $(\src\text{, } \dst\text{, } \ts_i\text{, } P_{i-1}\text{, } P_i\text{, } P_{i+1})$ and further divide it into two parts: $\callpp = (\src \| \dst \| \ts_i)$ and $\cdr = (P_{i-1}\|P_i\|P_{i+1})$, 
where in $\callpp$, $\src$ and $\dst$ are source and destination telephone numbers common to all providers in the call path, $\ts_i$(unique to $P_i$) is the time at which $P_i$ receives the call and $\cdr = (P_{i-1}\| P_i\|P_{i+1})$ are the previous hop, current hop and next hop respectively. Phone call setup takes time to traverse through the network, and we assume an upper-bound setup time $\tmax$. Any CDR pertinent to the same call will have a $\ts^*$ in the range of $[\ts_i - \tmax, \ts_i + \tmax]$.

To enable traceback, each \carrier{} $P_i$ contributes by sending command  $(\ccall, P_i, \callpp, \cdr)$ to the \RS.
The \RS{} will then add the record to their database $\DDD$. 
Later, if a party wishes to trace a certain call with $\callpp = (\src\|\dst\|\ts_i)$, they can send
the command  $\retreq$ to \RS{}, who will fetch all hops that have $\callpp = (\src\|\dst\|\ts^*)\text{, }\forall \ts^* \in [\ts_i - \tmax, \ts_i + \tmax]$.

Modeling a $\cdr_i$ for a \carrier{} $\provider_i$ as $(P_{i-1}\|P_i\|P_{i+1})$ allows $\provider_i$ attest to their upstream and downstream \carrier{}'s involvement in the call. This means that given a $\cdr_2$ from only $P_2$, we know the path $P_1 \rightarrow P_2 \rightarrow P_3$, so a traceback does not necessarily require records from $P_1$ and $P_3$. This design decision helps in partial deployment.

In the event of conflicting hops, for e.g., say $P_2$ submits $\cdr_2 = (P_1 \| P_2 \| P_3)$ indicating $P_1$ and $P_3$ as its previous and next hops. $P_1$ submits $\cdr_1 = (P_4 \| P_1 \| P_3)$. $P_3$ submits  $\cdr_3 = (P_1 \| P_3 \| P_6)$. In this case, an investigator cannot tell if $P_2$ is misbehaving or $P_1$ and $P_3$ are misbehaving. Therefore the investigator will go to each of these \carriers{} and have them show their corresponding call records to identify and punish the misbehaving \carrier{}(s).

This strawman solution trivially meets the functional requirements of the system, but none of the security requirements we described in Sec.~\ref{sec:requirements}. Indeed, since records are stored in the clear for \RS, no confidentiality is guaranteed to subscribers or \carriers{}. Furthermore, traceback could be done by any party with access to the records.

\myparagraph{Toward Record Confidentiality} 
To achieve record confidentiality, the first natural step is to encrypt the $\callpp$ and the $\cdr$. Assume all \carriers{} use a shared public key ($\pk$) to encrypt the $\cdr$ and $\callpp$ using an IND-CPA secure encryption scheme.
This means that \RS will store a set of ciphertexts, and hence cannot learn anything about the CDR content.
This guarantees the confidentiality of the records but unfortunately prevents the tracing process. Suppose an authorized party $P_j$ wants to trace call $\callpp$ = $(\src\|\dst\|\ts_j)$, they must send an encryption of $\callpp$ under $\pk$ to \RS. However, the \RS cannot find a matching record in the database, since the encryption scheme is not deterministic. 
Alternatively, the \RS sends all the ciphertexts in the database to $P_j$, and the latter decrypts each ciphertext until it finds the call records pertinent to their call. This approach is inefficient and loses call confidentiality for other calls. 
To solve this problem, we introduce a deterministic index to identify ciphertexts related to a given $\callpp$. Now upon receiving this index, the \RS can return exactly one ciphertext to the \carrier{}. We elaborate on this below.

\myparagraph{Adding Pseudorandom Labels to the Database}
To identify the correct ciphertexts, we index each entry with a label, $\clabel$, that can be computed only with the knowledge of $\callpp$. When a \carrier{} $P_i$ sends their contribution, they will send a pair ($\clabel_i$, $ctx_i$) to the \RS where $ctx_i$ is an encryption of $\cdr_i$ by $P_i$. Later, when a party  $P_j$ wants to trace a call $\callpp=(\src\|\dst\|\ts_j)$, they can use this information to compute $\clabel$, and \RS will be able to identify the $ctx$ that is indexed with $\clabel$. Note that $P_j$ will compute all  $\clabel^*$ for $\callpp = (\src\|\dst\|\ts^*)$ $\forall \ts^* \in [\ts - \tmax, \ts + \tmax]$ to retrieve all possible ciphertexts that belong to the call as specified earlier.

What function should we use to compute $\clabel$?
Perhaps the most natural candidate would be a hash function, i.e., $\clabel=$ $H(\callpp)$. 
However, this approach jeopardizes the confidentiality of the records once again. Indeed, anyone who gets access to the database $\DDD$ maintained by \RS can ``check'' if a certain call $(\src\|\dst\|\ts)$ took place by simply computing the hash of the call details and checking for that label in $\DDD$.
Adding a large nonce as input to the hash function i.e. $\clabel=$ $H(\callpp\|nonce)$ is not helpful since during trace the \carrier{}  will have to guess the nonce and this is infeasible in polynomial time. 

This attack suggests that the label should not be computed using a public function that anyone can compute. Pseudorandom Functions (PRF) are the perfect candidate. They are deterministic, just like hash functions, but they can be evaluated only with the knowledge of a key.
A label can be computed as $\clabel=$ $F_{k}(\callpp)$, where $k$ is a key known only by the \carriers{}  and $F$ is a PRF.
Hence, no one else, except carriers{}, can compute labels.
However, this solution is not robust in our threat model, where \RS could collude with \carriers{} .
Indeed, it would be sufficient for only one \carrier{}  to leak the PRF secret key $k$ to expose records.

We solve this problem using a cryptographic tool called an Oblivious PRF~\cite{casacuberta2022sok}  (OPRF). 
In an OPRF, the PRF is evaluated through a protocol between two parties: a server, who knows the key $k$, and a client, who knows the input $x$. At the end of the protocol, the client learns only the output of the PRF, while the server learns nothing.
In our system, we introduce a new party called the Traceback Authority (denoted \TA), which holds the secret key of the PRF and allows the \carriers{}  to evaluate the PRF to compute labels. The \TA, however, does not learn anything about the $\callpp$.

To contribute a record, $P_i$ will interact with the \TA to compute $\clabel_i$ from $\callpp$, and then compute $\idx_i=H(\clabel_i)$.
Next, $P_i$ will encrypt the $\cdr_i$ under $\pk$ into $ctx_i$ (as specified earlier) and submit $(\idx_i, ctx_i)$ to \RS.  The lookup index $\idx_i$ is a hash of the $\clabel_i$ so that if a record and/or the OPRF key is ever compromised, the $\clabel_i$ is not directly exposed. 

Traceback would work as follows: An authorized party $P_j$  who wants to trace  $\callpp$ first obtains the label $\clabel_j$ from \TA, then sends $\idx_j = H(\clabel_j)$ to \RS. \RS will use $\idx_j$ to identify and return the corresponding ciphertext.

However, there is still a problem:  
Recall that all ciphertexts are encrypted under the same key. Thus any \carrier{}  (potentially unauthorized) colluding with the \RS can potentially decrypt all ciphertexts trivially. 
Conversely, if each \carrier{}  encrypts its record using a unique key, the party attempting to perform a traceback will obtain ciphertexts under different keys, requiring \textit{all} \carriers{} to help with decryption. This defeats the purpose of the system.

\myparagraph{Towards Encrypting Records}\label{para:enc}
One potential solution is to have \carriers{} encrypt using the \TA's public key.  Thereafter, during the traceback, the \carrier{}  retrieves the ciphertexts from the \RS and interacts with the \TA to decrypt the ciphertexts.
While this is a viable solution, we want to formally enforce the following properties:  
\begin{enumerate}[itemsep=0pt, topsep=2pt]
    \item\noindent\underline{\textit{Knowledge of call:}} A party can trace a call only if they were part of the call i.e they {\em already know} the $\callpp$. The party must be a \carrier{} in the call path.
    \item\noindent\underline{\textit{Trace Authorization:}} A party can trace a call only if they were authorized by the \TA to trace that particular call.
\end{enumerate}

To this end we use an asymetric encryption scheme called ``witness encryption'' that allows carriers to encrypt the $\cdr$ such that only with the knowledge of the $\callpp$ and an authorization from the \TA can they decrypt the ciphertext. 
In witness encryption, the encryption key is a verification key $\vk$ for a signature scheme and an arbitrary string $\ell$ (chosen by the encryptor). A ciphertext can be decrypted only with the knowledge of the string $\ell$ and a signature on $\ell$ that verifies under $\vk$.
In our case, we replace $\ell$ with $\clabel$ to enforce property (1). We enforce property (2) by requiring a signature on $\clabel$ signed by the \TA. 

Now to contribute records, $P_i$ will compute $\clabel_i$, $\idx_i = H(\clabel_i)$ and encrypt $\cdr_i$ using $(\vk, \clabel_i)$ as the encryption key. Then $P_i$ will send $(\idx_i, ctx_i)$ to \RS. Using hash digests as indices instead of $\clabel$s further enforces property (1) above. Since the \RS is not part of the call path, it should not know the $\clabel$.

Once the ciphertexts are retrieved from the \RS, $P_i$ must request authorization from the \TA, in our case, a signature on $\clabel$. This construction ensures that a ciphertext related to a call can be decrypted only by someone who knows a valid signature on the corresponding $\clabel$.

\myparagraph{Adding Carrier Anonymity and Accountability} Recall that to contribute, each \carrier{} sends authenticated encrypted records signed under their unique public key to \RS. The \RS can map their contributions to their identity, potentially learning trends about their activities. On the other hand, we cannot simply have the \carrier{} submit their records anonymously since we still need to hold them accountable for malformed or falsified contributions. 
 
To protect carriers' business privacy but hold them accountable, we replace the regular signature scheme (used for authentication) with an anonymous group signature scheme. 

Group signatures are anonymous signatures that can be validated on behalf of a group -- in our case, the group of all carriers. More importantly, we choose group signatures instead of primitives like ring signatures because group signatures are efficient and allow us to trace traitors. Here, we use group signatures only for authenticating contribution requests and not for the witness encryption scheme. 

In our system, the \TA plays the role of the group manager and adds carriers to the group by assigning them group secret keys. Moreover, the \TA is also responsible for the deanonymization of the group signatures in case any of the carriers submit bad requests. 

\myparagraph{Network Layer Anonymity}
The group signature scheme for authenticating contribution requests guarantees anonymity at the application layer. Unfortunately, network features like IP addresses may still identify \carriers. There are a number of solutions to this problem that are orthogonal to \sysname, including proxy services like commercial VPNs. A \carrier{} may still be concerned that IP traffic volume might leak information about call volumes to the proxy.  Providers can address this issue by splitting their traffic across multiple proxy services and/or  transmit redundant or invalid records as  cover traffic.

\subsection{Threat Model and Resiliency}
\label{sec:threat-model}
\projectname mandates the security requirements outlined in Sec~\ref{sec:requirements}. 
The \projectname system includes two entities besides providers: the \RS and the \TA. 
We assume that the \RS and the \TA do not collude and only one of the two entities may be malicious. We also allow collusion between \carriers and the corrupt entity. We show that even when the \RS is malicious and is colluding with providers, none of the security requirements are violated. On the other hand, when the \TA is corrupt, \projectname cannot guarantee record accountability. 

We can improve the trustworthiness of the \TA with several orthogonal techniques, described below. All of these options are feasible, but they add complexity and cost.

\myparagraph{Splitting responsibilities} In the architecture of \projectname, the \TA is responsible for group management, label generation, and authorizing traceback.
Assigning these jobs to different entities will limit the damage should one be compromised, and our prototype actually already implements them independently. 

\myparagraph{Distributing Trust}
Each of the operations can also be split among multiple entities using existing multiparty computation schemes, including threshold constructions of OPRFs\cite{jarecki2017toppss,baum2020pesto}, group signatures\cite{camenisch2020short,ghadafi2014efficient}, and BLS signatures
for use in Witness Encryption.

 \subsection{Frequently Asked Questions}
\label{sec:scope}
In this section, we address some of the frequently asked questions that we have encountered.

\myparagraph{If \StirSha automates traceback, why go through all this trouble?}
The Public Switched Telephone Network (PSTN) is heterogeneous. Legacy infrastructure drops \StirSha signatures along the call path, making it ineffective for tracing call origins \cite{bercu23}. 
Call requests are signed by providers using a JSON Web Token in the SIP INVITE message with an $\mathtt{x5u}$ field pointing to the signing certificate. Malicious providers can exploit this by setting $\mathtt{x5u}$ to a timing-out link, increasing latency and forcing call transmission, posing a challenge for providers. Traceback attempts using such signatures reach dead ends, so manual processes are still needed. Additionally, the deployment of \StirSha remains limited. According to the Robocall Mitigation Database in the US (Feb 7, 2024) \cite{robomitidb}, among 7,109 providers, only 39.94\% have fully implemented it, 23.96\% are partially implemented, and 36.11\% have no or unknown implementation status. We clarify that \sysname does not authenticate caller ID or block robocalls in real-time. Instead, it is a central repository of encrypted CDRs for traceback purposes. 

\myparagraph{Why can't we just put traceback info in headers}
Implementing traceback information in headers faces the same hurdles as S/S.

\myparagraph{Why develop a new protocol instead of automating the manual tasks done by groups in ITG?} The ITG currently maintains a semi-automated traceback system that sends notifications to \carriers{} who are mandated to respond within 24 hours. Automating the current traceback tasks will require all providers to implement a traceback API that integrates with the ITG systems. While this automates the process, the gain on ``traceback throughput'', the number of computable traceback requests per month, remains low as the process is still serial and involves \carriers{} active participation for every traceback request. Tracebacks would run into dead-ends if a single \carrier{} does not cooperate, their portal goes down, responds with misleading information, or partial deployment. 

For \carriers{}, this alternative adds an additional cost of maintaining an inbound traceback system. An adversary could exploit vulnerable API implementations of this mandate to access affected \carriers{}' sensitive data. Note that compromising a carrier's API server exposes its peers and  network trends as well as subscriber call history, and well-funded companies who use industry best-practices are regularly breached. 
We designed \sysname as a centralized distributed system to alleviate serial traceback lookups, enabling providers to be passive entities rather than active for traceback computation. This design choice not only enhances the traceback throughput but also centralizes security management from several thousand providers to two organizations responsible for overseeing \sysname. Moreover, if any single \sysname entity is compromised, no plaintext data is leaked.

\myparagraph{Why require participation from all providers if only the originating provider's records are needed?} If only originating providers submit records, traceback fails because malicious carriers likely won't comply. Limiting submissions to the originator and second hop is infeasible because providers cannot determine their sequence in the call path. Thus, requiring all providers to submit records becomes essential to trace back and identify people facilitating bad calls. Having the ability to construct the full call path has added advantages such as trace-forwards to debug call routing or blocking errors.

\myparagraph{Who will operate the \TA and \RS?}
The telephone network is the \emph{ideal} environment for \sysname because regulators like the FCC already designate trusted third parties that provide singleton functions. Examples include toll-free numbering, local number portability (LNP) databases, \StirSha certificate authority governance, and the current traceback clearinghouse, ITG. The FCC periodically solicits applications to serve as these entities and a fair and competitive process among several for-profit enterprises follows. The selected entities can then charge reasonable fees for the service they provide.

For \sysname, there are two entities to recruit. The \RS roughly corresponds to an entity like the LNP databases, and many vendors have the technical ability to serve in this role. 
The TA performs functions similar to the current ITG, like registering providers to their system and determining if a traceback query is appropriate, so modifying that existing role would provide a straightforward on-ramp to deployment.

\myparagraph{What if \carriers{} refuse to participate?} \sysname must be mandated to be effective. Regulation can compel \carriers{} to participate or have their network access revoked. Unlike \StirSha, \sysname is compatible with all network technologies in use.

\section{System and Protocol Design} 
\label{sec:protocol-design}
In this section, we discuss \projectname's system architecture and detailed protocol.
\begin{figure}[t]
  \centering
  \includegraphics[width=3in]{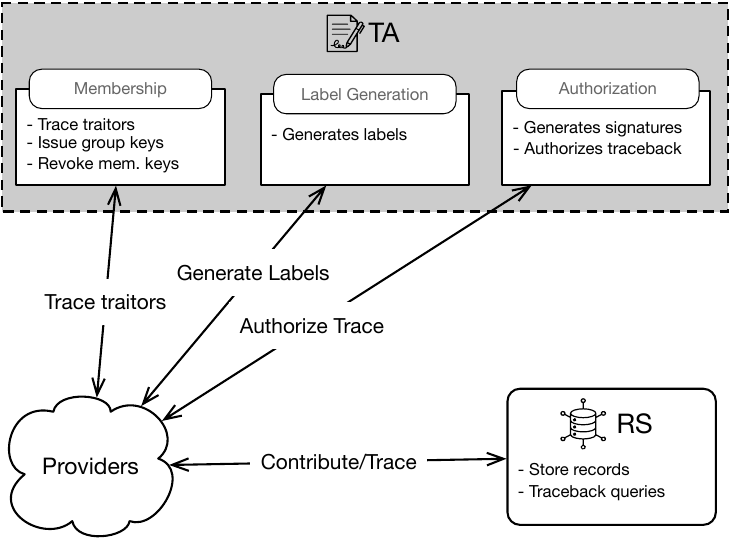}
  \caption{\projectname{} facilitates efficient and rapid traceback through collaborative efforts. \Carriers{} engage with the \TA to secure membership, create labels, and acquire trace authorizations. Additionally, carriers send(receive) encrypted records to(from) the \RS}
  \label{fig:system-diagram2}
\end{figure}

\subsection{\projectname Architecture}
\label{sec:system-design}
Figure \ref{fig:system-diagram2} shows a high-level diagram of \projectname's system architecture.

\begin{table}[t]
    \centering
    \caption{Notation used in the protocol}
    \footnotesize
    \begin{tabular}{|l|l|}
    \hline
        \bf Notation & \bf Description  \\\hline
        \RS & Record Store \\
        \TA & Traceback Authority \\\hline
        $P_i$ &  A carrier \\
        $\src$ & Telephone number of call initiator \\
        $\dst$ & Telephone number of call recipient \\
        $\ts_i$ & Time call reached \carrier{} $P_i$ \\
        $\callpp$ & Defined as $\src\|\dst\|\ts$ \\
        $\cdr_i$ & Hop submitted by $P_i$ defined  as $P_{i-1} \| P_i \| P_{i+1}$ \\
        $\clabel$ & A label for a call\\
    \hline
    \end{tabular}
    \normalsize
    \label{tab:notations}
\end{table}

\noindent 
There are three kinds of entities in our system: 

\myparagraph{Carriers $\mathbf{P_i}$} 
A carrier $P_i$ receives calls from either the source ($\src$) or from a previous carrier ($P_{i-1}$) and forwards the call to either the destination ($\dst$) or the next carrier $P_{i+1}$. A call is identified by its $\callpp = (\src\|\dst\|\ts)$ where $\ts$ is the time at which the call reaches the carrier.

\myparagraph{Record Store (\RS)}
The \RS{}  maintains a database that stores {\em encryptions} of the hops associated with a call.  Recall that a  hop is a tuple $\cdr := (P_{i-1}\|P_i\|P_{i+1})$.

\myparagraph{Traceback Authority (\TA{})}
The \TA  has the following functions: 
\begin{itemize}[itemsep=0pt, topsep=2pt]
    \item\noindent\underline{\textit{Authorizing Trace Requests}}: The \TA provides the signatures that enable a carrier to decrypt records retrieved from the \RS. These signatures are computed on the call labels.

    \item\noindent\underline{\textit{Managing \Carriers'  Anonymous Authentication}}:
    The \TA{} manages the group signatures for the carriers. This consists of adding legitimate \carriers{} to the group and providing them with credential to sign on behalf of the group. The \TA is responsible for accountability, it can deanonymize signatures in case of misbehavior and hold the corresponding entity responsible.
    
    \item\noindent\underline{\textit{Generating Pseudorandom Labels}}: The \TA interacts with carriers to compute $\clabel$s.

\end{itemize}

\begin{figure}[t]
  \centering
  \includegraphics[width=3.25in]{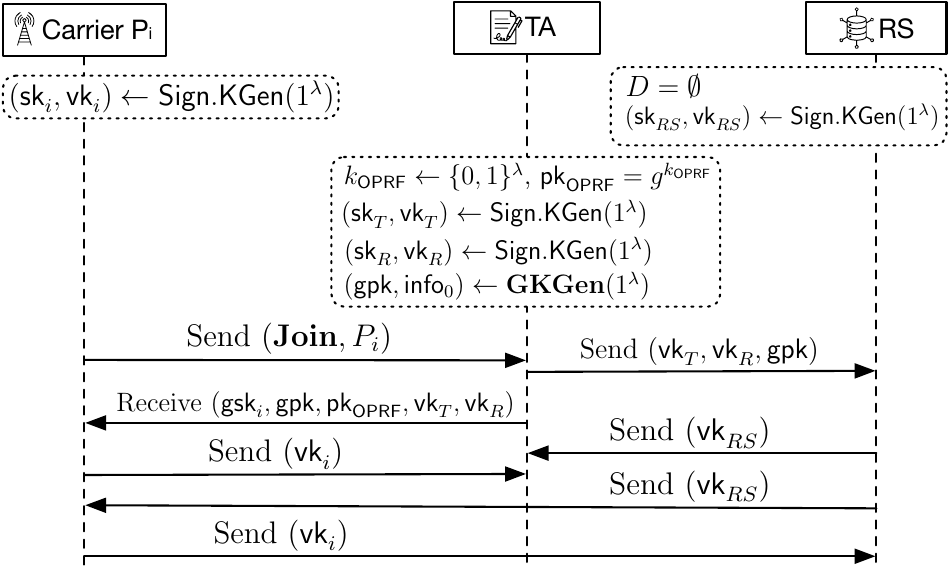}
  \caption{During setup protocol, \TA~ generates and announces public parameters. \Carriers{} request group membership and are assigned member secret key upon acceptance by the \TA}
  \label{fig:protocol-registration}
\end{figure}

\subsection{Protocol Overview}
\projectname~ consists of four protocols:  \emph{\protosetup}, \emph{\protocontribute}, \emph{\prototrace}, and \emph{\protoopen}, which we describe in detail below.

\myparagraph{Cryptographic Primitives} 
As discussed in Sec. \ref{subsec:protocol-prelims},
\sysname uses a witness encryption scheme for signatures (WES)\cite{wes,mcfly}, a group signature scheme\cite{bootle2016foundations}, an oblivious PRF protocol\cite{jarecki2017toppss}, signature schemes, and a hash function. The notations used in our protocols are described in Table~\ref{tab:notations}. \ifisFullVersion We present the protocol in detail in Appendix~\ref{prot:details}. \fi Below we present an overview of the protocol.

\myparagraph{\protosetup ~Protocol}
The \protosetup protocol is described in Fig.~\ref{fig:protocol-registration}.
This protocol is run by the \TA and carriers to set up their keys.

{\bf TA Setup:}
The \TA sets up (1) the group with a group master key and secret key. (2) the PRF key for the oblivious PRF and announces a public key (denoted $\pk_\oprf$) corresponding to the PRF key. (3) two signature key pairs $(\sk_T, \vk_T)$ and $(\sk_R, \vk_R)$ and announces $\vk_T$ and $\vk_R$ to all entities. Here, signatures using $\sk_T$ will be used to decrypt the witness encryption ciphertexts, and $\sk_R$ will be used to authorize trace requests.

{\bf Carrier $\mathbf{P_i}$ Setup:}
Each \carrier{} (denoted $P_i$) joins the system by first interacting with the \TA to get a distinct group signing key $\gsk_i$, which they can use to sign anonymously on behalf of the group. They generate a regular signing key pair $(\sk_i, \vk_i)$ for authenticated communication with the \TA and \RS (during trace). 

Finally the \RS initializes a database $\DDD$ and sets up a signature key pair $(\sk_{\RS}, \vk_{\RS})$ and announces $\vk_{\RS}$

\begin{figure}[t]
  \centering
  \includegraphics[width=3.3in]{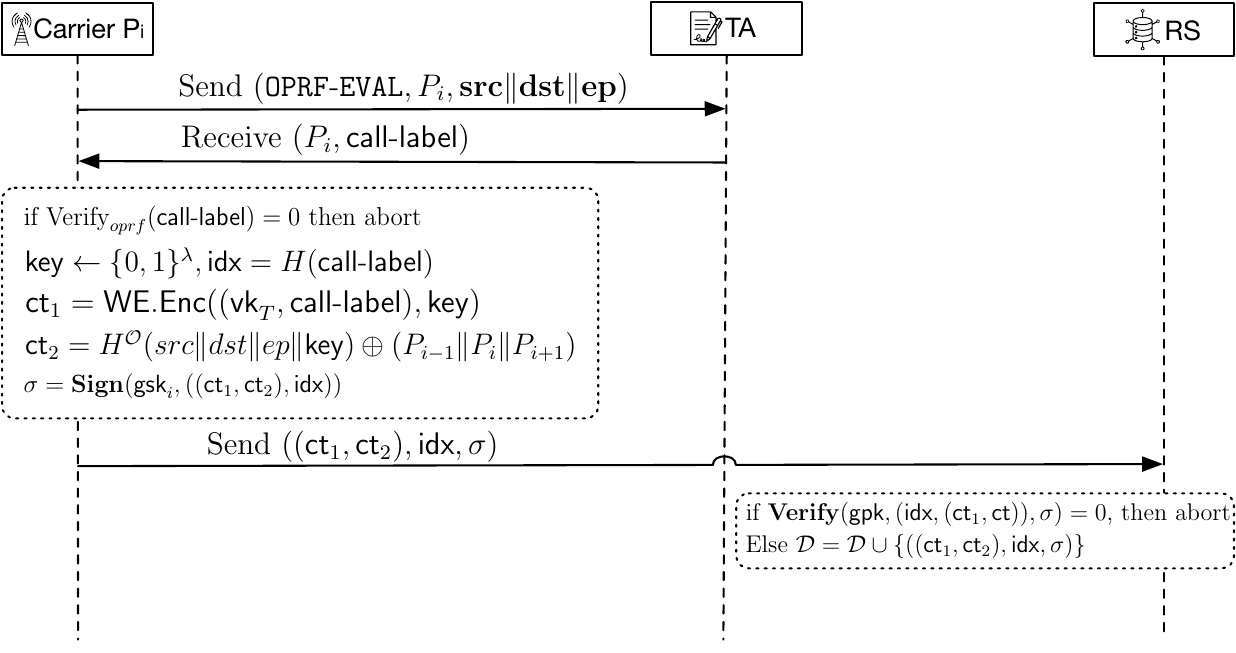}
  \caption{In contribution protocol, \carrier{} submits ciphertexts, compliant with the protocol, to the Record Store}
  \label{fig:protocol-contribution}
\end{figure}

\myparagraph{\protocontribute ~Protocol}
\Carriers{}
record the $\cdr$s with the \RS using the \protocontribute protocol. We consider the case of submitting a single $\cdr$ to the \RS{} in Fig. \ref{fig:protocol-contribution}. Each \carrier{} in the call path parses the CDR into $\callpp=(\src\|\dst\|\ts)$  and $\cdr = P_{i-1}\|P_i\|P_{i+1}$ as defined in Sec~\ref{sec:approach}. To contribute call records, the \carrier{} $P_i$ anonymously submits a witness encryption of the message $\cdr = (P_{i-1}\|P_i\|P_{i+1})$ to \RS, with a pseudorandom label associated with it and a group signature for authentication.
The ciphertext and the label leak no information about the call thus providing confidentiality of the call, and the group signature leaks no information about the \carrier{} sending this information, thus providing anonymity to the carrier.
We elaborate on how the label, the encryption, and the signature are computed below.

The \protocontribute protocol consists of two phases: the label generation phase and the submission phase. In the label generation phase,  the label is computed with the help of the \TA, using an oblivious PRF protocol.  The \TA acts as a server and holds a PRF key, and the \carrier{} acts as a client with the input. 
We assume that each $\ts$ is associated with an epoch, $\ts$ truncated to nearest centisecond, denoted $\ep$.
The \carrier{} uses $\callpp = (\src\|\dst\|ep)$ as input and the output of the protocol (denoted  $\clabel$) is learned only by the \carrier{}. We note that with the help of $\pk_\oprf$, the \carrier{} can compute an efficient pairing check to verify that the output received from the \TA is indeed correct, and  that the PRF key $k$ was used to compute the $\clabel$ \ifisFullVersion as detailed in Fig.~\ref{fig:oprf-scheme}\fi.
The \carrier{} then computes $\idx = H(\clabel)$. Recall that $\idx$ is used instead of $\clabel$ to index the ciphertexts to prevent the \TA from trivially decrypting all ciphertexts in the case that the database of ciphertexts and $\clabel$s  are leaked. 

In the submission phase, the \carrier{}  prepares the encryption as follows. $P_i$ samples a $\lambda$-bit uniform $\cdrkey$ from $\bin^\lambda$ and encrypts $\cdrkey$ with the WES scheme using  $(\vk_T, \clabel)$ as the encryption key to get a ciphertext $ct_1$. The WES scheme ensures that the ciphertext can only be decrypted using a signature on $\clabel$ signed using $\sk_T$ by the \TA. $P_i$ further computes $\ct_2 = \cdr \oplus H(\callpp\| \cdrkey)$, where $H$ is modeled as a random oracle. 
Note that we require $\callpp$ as input to the hash function to extract the $\callpp$ in the proof of security. 
Finally, the \carrier{} signs the message $((ct_1, ct_2), \idx)$ using the group signature scheme, obtains $\sigma$, and sends the resulting tuple $((ct_1, ct_2), \idx, \sigma)$ to the \RS. Upon receiving a submission request, \RS validates the group signature $\sigma$ and stores the tuple in the database; if verification fails, the request is dropped.

\begin{figure*}[t]
  \centering
  \includegraphics[width=0.9\textwidth]{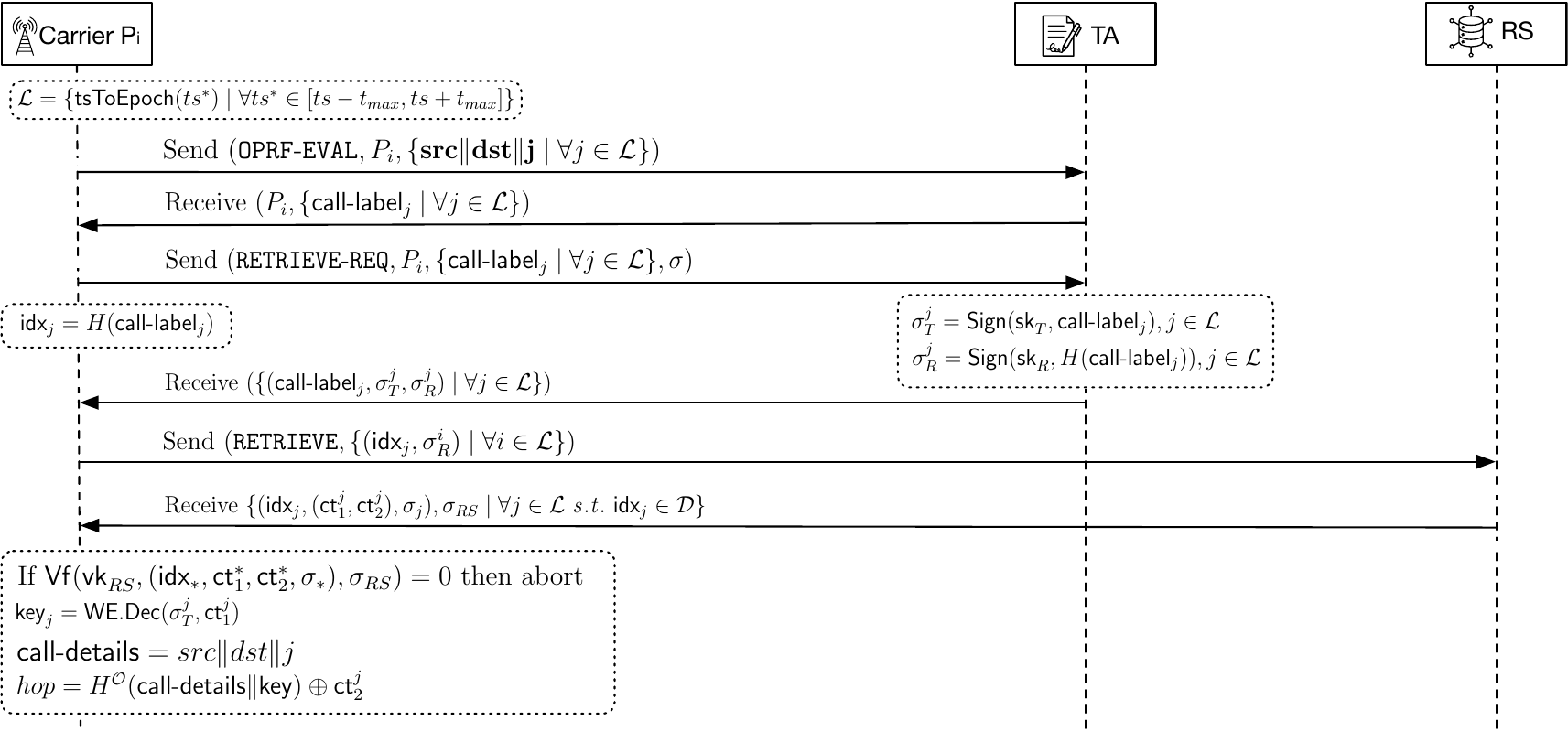}
  \caption{In the trace protocol, \carrier{}s obtain labels, ciphertexts, and decryption authorization signatures for a call from \TA. Here, the bold text shows $src\|dst\|j$ is not sent in plaintext but blinded as in OPRF }
  \label{fig:protocol-trace}
\end{figure*}

\myparagraph{\prototrace Protocol}\label{proto:trace}
To trace a call, the \carrier{} must retrieve all the $\cdr$s corresponding to a call from the \RS. Fig.~\ref{fig:protocol-trace} illustrates the sequence of events in the trace protocol. Initially, the \carrier{} needs to obtain the labels corresponding to these calls. For this purpose, the \carrier{} computes the labels with the assistance of the \TA using the OPRF protocol described above. We assume an upper limit $\tmax$ on the duration of a call setup from the source to the destination. The \carrier{} computes epochs $\ep$ corresponding to the timestamps $\ts* \in [\ts - \tmax, \ts + \tmax]$ and computes the labels associated with the $\callpp$ for each of these epochs. 

Before we describe how the \carrier{} gets the full trace, we note that a malicious \carrier{} colluding with the \RS could potentially compute labels on arbitrary $\callpp$ and check with \RS if such labels exist in the store, revealing information about the existence of a call between a certain initiator and recipient. If the \carrier{} attempts to perform this attack for a specific initiator and recipient, we call it a ``targeted attack''. If the adversary's goal is to map the network by trying to compute labels on arbitrary $\callpp$, we refer to such attacks as ``grinding attacks''. 
To mitigate the grinding attack, we implement rate-limiting on the number of requests a \carrier{} can make. For this purpose, when the \carrier{} interacts with the \TA to compute a $\clabel$, the \TA maintains a count of requests made by a \carrier{} and will not authorize further requests if a certain limit is exceeded. To give authorization on this request, the \TA will sign  $\idx = H(\clabel)$ using $\sk_{R}$, and this signature $\sigma_R$ serves as an authorization for traceback. 
Moreover, consider the case that the \TA is malicious (and the \RS is honest) and is colluding with a \carrier{}, then they can easily mount the grinding attack described above by computing arbitrary labels and requesting the corresponding $\idx$ from the \RS. To mitigate this, we will also require the \RS to implement rate-limiting on trace requests. This will ensure that a \carrier{} colluding with a \TA cannot mount grinding attacks. 

The \carrier{} requests the \RS for the ciphertexts corresponding to the $\idx$. The \RS first checks that the signature $\sigma_R$ is verifiable using $\vk_R$. It rejects the request if this is not the case. 
The \RS identifies the ciphertexts corresponding to the $\idx$ in $\DDD$ and sends $\ct_1, \ct_2, \idx, \sigma_{\RS}$, where $\sigma_{\RS} = \sign(\sk_{RS}, (\ct_1, \ct_2, \idx))$.
The \carrier{} then asks the \TA for a signature on the $\clabel$ and uses these signatures to decrypt the ciphertexts and compute the $\cdr$s.

\myparagraph{\protoopen ~Protocol}
Some \carriers{} may provide wrong hops to frame others. Since the encrypted hops are anonymously submitted to the \RS, we need a mechanism to catch the malicious \carrier{}s. Our group signature protocol allows the \TA to {\it open} any group signature and reveal the \carrier{} that signed a message. Thus, if a trace seems malformed, the \carrier{} submits all ciphertexts, hops, and signatures for the call retrieved from the \RS and sends it to the \TA. The \TA{} runs a $\validate$  function that outputs the set of faulty hops. The \GM{} then identifies the signatures that correspond to these hops and deanonymizes them to return the set of \carriers{} that submitted malformed/wrong hops.

\begin{figure*}
    \centering
    \begin{tcolorbox}[
    standard jigsaw,
    opacityback=0]
    \begin{flushleft}
    \smallskip\noindent\underline{$\validate(\{\callpp_j,\clabel_j, (ct_0^j, ct_1^j), \sigma_j,\sigma_T, \cdr_j\}_{\cdr_j \in \CCC_\trace})$}: The validation algorithm proceeds as follows:
    
    \begin{enumerate}
        \item Verify that $\clabel_j$ corresponds to the call that is to be validated using $\callpp$.
        \item Verify that $\cdr_j$ can be computed from $(ct_0^j, ct_1^j)$ using $\clabel_j$. 
        \item Each \( \cdr_j \) is defined as \( \provider_{i-1} \,||\, \provider_{i} \,||\, \provider_{i+1} \). Create a directed multigraph \( \mathcal{G} \) such that: 
        \( \mathcal{G}.V = \{ P_k \mid P_k \in \cdr_j, \forall \cdr_j \in \CCC_\trace \} \) and 
        $\mathcal{G}.E = \{\{\provider_{i-1} \rightarrow \provider_{i},
                \provider_{i} \rightarrow \provider_{i+1}
                \mid \forall \cdr_j \in \CCC_\trace\}\}$.  We use $\{\{\cdot\cdot\cdot\}\}$ to denote that $\mathcal{G}.E$ is a multi-set.

        \item For each $v \in \mathcal{G}.V$:
            \begin{enumerate}
                \item If $\mathcal{G}.\mathsf{deg_{in}}(v) = 0$, add $v$ to $\mathcal{O}$: \emph{list of \carriers{} that claim to be originators}.
                \item If $\mathcal{G}.\mathsf{deg_{in}}(v) > 0$ and $\mathcal{G}.\mathsf{deg_{out}}(v) = 0$, add $v$ to $\mathcal{D}$: \emph{list of \carriers{} that claim to be terminating}.
                \item Else add $v$ to $\mathcal{T}$: \emph{list of \carriers{} that claim to be transit}.
            \end{enumerate}
        \item If $|\mathcal{O}| = 1 \text{ and } \mathsf{deg_{out}}(v) > 0$, set $\mathcal{P}_O = v \mid v \in \mathcal{O}$ else do:
            \begin{enumerate}
                \item For $v \in \mathcal{O}$: if $\mathcal{G}.\mathsf{deg_{out}}(v) = 2$ add $v$ to $\mathcal{O}_T$ (\emph{possibly true origins}) else add $v$ to $\mathcal{O}_F$ (\emph{possibly faulty or missing attestation})
                \item If $|\mathcal{O}_T| = 1$ set $\mathcal{P}_O = u \mid u \in \mathcal{O}_T$: \emph{(true origin)}
                \item $\mathcal{F}_O = \mathcal{F}_O \cup \mathcal{O}_F$: \emph{may require investigation from output graph}
            \end{enumerate}
        \item If $|\mathcal{D}| = 1 \text{ and } \mathsf{deg_{in}}(v) > 0$, set $\mathcal{P}_T = v \mid v \in \mathcal{D}$ else do:
            \begin{enumerate}
                \item For $v \in \mathcal{D}$: if $\mathcal{G}.\mathsf{deg_{in}}(v) = 2$ add $v$ to $\mathcal{D}_T$ else add $v$ to $\mathcal{D}_F$
                \item If $|\mathcal{D}_T| = 1$ set $\mathcal{P}_T = u \mid u \in \mathcal{D}_T$ \emph{(true terminator)}
                \item $\mathcal{F}_D = \mathcal{F}_D \cup \mathcal{D}_F$
            \end{enumerate}
        \item For $v \in \mathcal{T}$: if $\mathcal{G}.\mathsf{deg_{in}}(v) \notin \{1,2\}$ and $G.\mathsf{deg_{out}}(v) \notin \{1,2\}$ add $v$ to $\mathcal{F}_T$.
        \item If $\mathcal{G}$ is weakly connected, then output $\mathcal{G}.\text{shortestPath}(\mathcal{P}_O, \mathcal{P}_T)$ else $\mathcal{G}.\text{subgraphs}()$
        \item Output ($\mathcal{F}_O$, $\mathcal{F}_T$, $\mathcal{F}_D$), ($\mathcal{P}_O$, $\mathcal{O}$), ($\mathcal{P}_T$, $\mathcal{D}$), ($\mathcal{T}$)
    \end{enumerate}
    \end{flushleft}
    
    \end{tcolorbox}
\caption{The Validate algorithm analyzes deviations from the ideal scenario to determine the call path and faulty sets}
    \label{fig:validate-hops}
\end{figure*}

\subsection{Traceback Validation}\label{sec:traceback-validation}
Once a \carrier{} has $\cdr$s, they must assemble the hops into the complete path. We call this step ``Traceback Validation'' because it will identify the correct path or detect inconsistencies or missing records that should be manually investigated.

We validate a traceback by inserting decrypted records into a directed multi-graph.
In a multi-graph, two nodes can be connected by multiple edges. 
Each record is a graph with three nodes: $\provider_{i-1}\text{, }\provider_{i}$ and $\provider_{i+1}$; and two edges: $\provider_{i-1} \rightarrow \provider_{i}$, and $\provider_{i} \rightarrow \provider_{i+1}$. 
A traceback is the multi-graph union of such individual sub-graphs(hops). 

\begin{figure}
    \centering
    \includegraphics[width=\columnwidth]{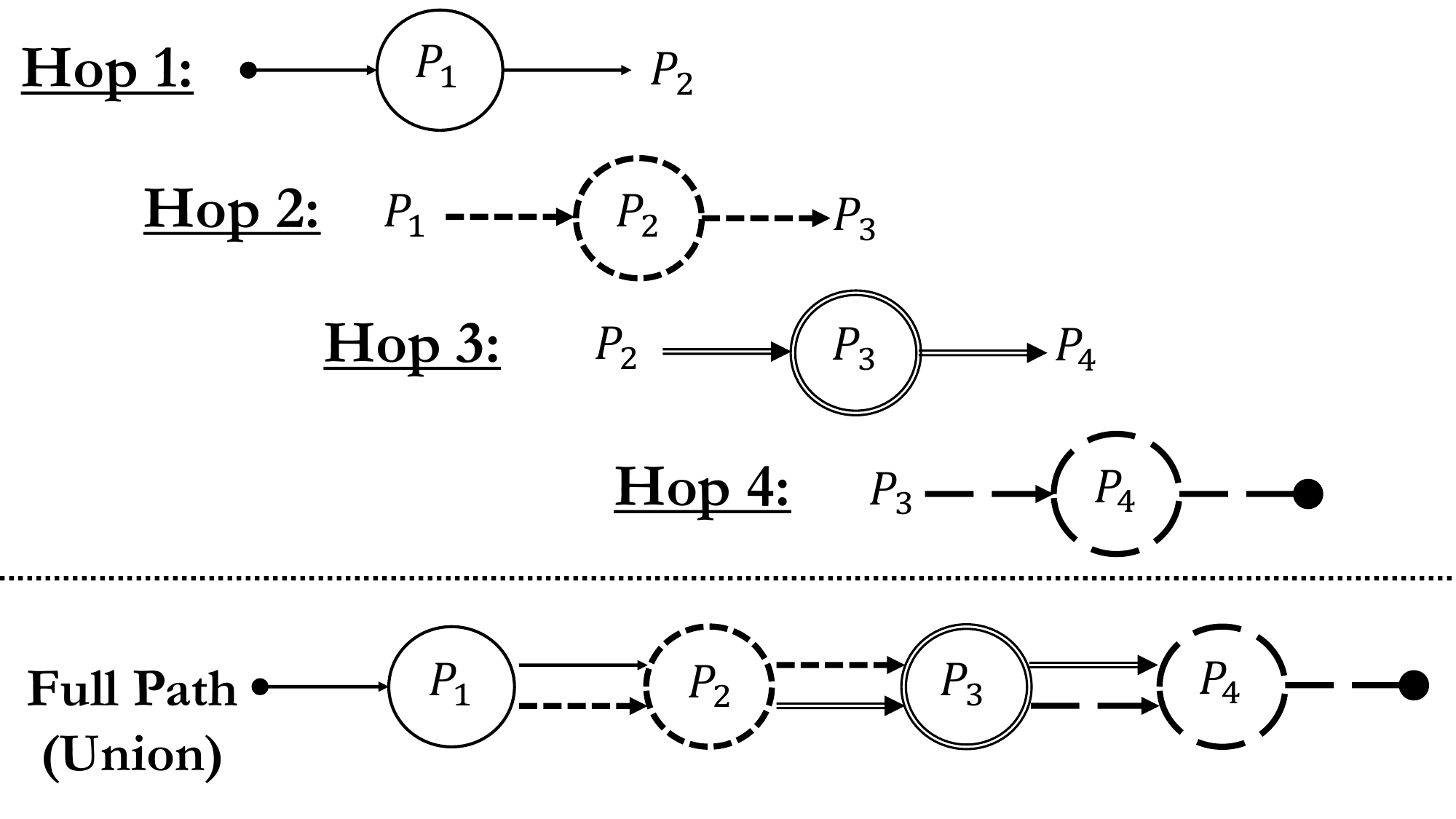}
    \caption{The full call path is a union of subgraphs from $\cdr$s}
    \label{fig:hops-to-full-path}
\end{figure}

\myparagraph{Ideal Scenario} Each $\provider_i$ contributes a valid record for a given call. Figure~\ref{fig:hops-to-full-path}(Full Path) visualizes this ideal scenario. Let $\text{deg}_{in}$ and $\text{deg}_{out}$ denotes in and out degrees respectively. The originating \carrier{} $\provider_1$ has $\text{deg}_{in}=0$ and $\text{deg}_{out}=2$. Here, one edge of $\text{deg}_{out}$ (denoted by solid line) indicates its assertion to $P_2$ while the other edge (denoted by dashed-line) indicates $P_2$'s assertion to $P_1$. The terminating \carrier{} $\provider_4$ has $\text{deg}_{in}=2$ and $\text{deg}_{out}=0$ for reasons symmetrical to one provided for the originating provider. Finally, each transit \carrier{} ($\provider_2$ and $\provider_3$) has $\text{deg}_{in}=2$ and $\text{deg}_{out}=2$.
Any deviation from the ideal case helps us determine ``faulty hops''---hops that are conflicting.

\myparagraph{Determining Faulty Hops} We detect faulty hops by checking four properties formulated from the ideal scenario:

\smallspace\noindent\underline{\it Origin invariant}: A call can have only one originator. This property holds if there is exactly one node in the directed multi-graph with $\text{deg}_{in}=0$ and $\text{deg}_{out}\in \{1, 2\}$. Otherwise, the originating record is missing, or some $\provider_i$ submitted malformed records. Note that $\text{deg}_{out} = 1$ does not necessarily imply a malicious originator since true originators will still have $\text{deg}_{out} = 1$ if there is a missing record from their downstream \carrier{}. If there is more than one node with $\text{deg}_{out} = 2$, then there are conflicting originators; in this case, we construct different call paths for each originator.

\smallspace\noindent\underline{\it Terminating invariant}: A call can have only one terminating provider. This property holds if there is exactly one node with $\text{deg}_{in} \in \{1, 2\}$ and $\text{deg}_{out} = 0$. Otherwise, the terminating record is missing, or a $\provider_i$ submitted malformed records. This is symmetric to the origin invariant except that values for $\text{deg}_{in}$ and $\text{deg}_{out}$ are swapped.

\smallspace\noindent\underline{\it Transit invariant}: This property identifies all transit providers and validates for nodes having $\text{deg}_{in} \in \{1, 2\}$ and $\text{deg}_{out} \in \{1, 2\}$.

\smallspace\noindent\underline{\it Connectivity invariant}: This property determines if the full call path can be recovered. It holds if there is a path between every pair of vertices in the traceback graph.

Figure \ref{fig:validate-hops} presents the detailed algorithm for determining the faulty hops from the decrypted records and the ability to reconstruct the call path.

\myparagraph{Traceback Robustness and Partial Deployment} 
In a scenario where all parties are honest, we derive all the benefits of the scheme. Importantly, even in partial deployment scenarios where only certain providers submit their records, our scheme can identify the call originator under certain conditions. For example, if the originating provider is the sole contributor for a call, the system can still successfully identify the call origin without the contributions from any other provider in the call path.
If the second hop provider participates, our system can identify the origin even if no other provider participates. If there are ever conflicting origin claims, we initiate a manual investigation to identify the source and punish the dishonest party. If any other intermediate party participates honestly, we can still reduce manual traceback time. With the call path recovery algorithm, there may be cases where we can precisely identify the bad actor. However, this is an ``added bonus'' and not the main goal of the scheme.
The tracing algorithm relies on a best-effort strategy, and we believe that societal incentives, including civil or criminal liability, will motivate the entities to behave honestly. In Sec. \ref{sec:traceback}, we present an evaluation of \sysname in partial deployment.

\subsection{Security of ~\projectname}
In this section we provide informal arguments that \projectname{}  achieves the security properties outlined in Sec~\ref{sec:requirements}. The formal proof in the UC framework is in \ifisFullVersion Appendix~\ref{sec:ucproof} \else our extended technical report \cite{jager_extended} \fi.

\begin{theorem} {\em [Informal]} Assuming the CPA security of the witness-encryption scheme, the unforgeability of the signature scheme, the security of the group signature scheme, the security of the OPRF protocol, and secure hash functions, \projectname{} achieves record confidentiality, the privacy of individual caller, blinds network trends and \carrier{} associations. Moreover if the \TA is honest, \projectname additionally achieves \carrier{} accountability. 
\end{theorem}

\myparagraph{Trace Authorization} Since a \carrier{} requires a signature from  the \TA to decrypt the ciphertexts, and a \carrier{} needs to know the $\callpp$ to request this signature, only authorized parties can perform a trace successfully.

\myparagraph{Call Confidentiality} Recall that the $\src, \dst$ information of the call is used only to compute the label. Since the computation of the label is through an OPRF, we guarantee that the \TA does not learn the $\src$ and $\dst$ of the call. Since the OPRF output is pseudorandom, the label by itself will also not reveal any information about the caller and the callee of the call.

\myparagraph{Trade Secret Protection} Recall that each $\cdr$ is encrypted and stored at the $\RS$. Decrypting the encrypted $\cdr$ requires knowledge of the labels and a signature on the $\clabel$, information accessible only to entities involved in the call or those accurately guessing the source, destination, and call time. The unforgeability property of the signature scheme ensures that an entity cannot forge a signature on behalf of the \TA{}, preventing unauthorized decryption of the $\cdr$s. Additionally, the CPA security of the WES scheme safeguards the contents of these ciphertexts.
As previously described, a malicious \carrier{} might attempt to guess arbitrary call details, create corresponding labels, and decrypt records stored at the $\RS$, i.e. they try to mount a grinding attack. To mitigate this risk, we implement rate-limiting on such requests by having the \TA{} restrict the number of authorizations granted to each \carrier{}. 

\myparagraph{Record integrity} Since all contributed records are signed using a group signature, and the \RS verifies this signature before adding the record to the database, we ensure that only authorized users can contribute records. 

\myparagraph{Record Accountability} We achieve \carrier{} anonymity since each submission to the $\RS$ does not include any identifier of the \carrier{}. They are instead signed using the group signature scheme, ensuring that the \carrier{} is anonymous within the group. When a \carrier{} is misbehaving (e.g. by sending a malformed $\cdr$) the group signature can be opened by \TA thus revealing the \carrier{} that signed the submitted $\cdr$. 
We note that if the \TA is malicious they may just not reveal any identity and accountability may not be guaranteed. But even if the \TA is malicious they cannot frame an honest \carrier{} as the sender of the record.

\section{Implementation}

This section describes the prototype implementation of \sysname and how we obtained CDR data to evaluate it.

\subsection{Prototype Implementation}

We describe a prototype implementation for each component of \projectname, which enables us to evaluate its performance.

\myparagraph{Traceback Authority} We implement the \TA as an HTTP server that uses BLS Signatures \cite{boneh2019bls} to compute authorization signatures. For this function, we exposed an endpoint for authorizing trace requests. We set up our group signature scheme using short group signatures~\cite{boneh2004short} implemented by IBM's \texttt{libgroupsig}~\cite{libgroupsig} and exposed an endpoint for opening signatures. Finally, we used the \texttt{ristretto255} elliptic curve (\texttt{oblivious} Python package) for our OPRF protocol~\cite{burns2017ec}. The \TA exposes an API endpoint for label generation.

\myparagraph{Record Store} The \RS{} is an HTTP server with a database for storing records. We use the \texttt{Click-House} columnar database. The \RS{} exposes endpoints for contribution and traceback queries.

\myparagraph{Carrier} We implement a carrier as a process that runs the protocol and interacts with other components in the system. For performance, we implemented the witness encryption scheme in C++ using the \texttt{BLS381} elliptic curve  library\cite{chia_bls_signatures} and wrote Python bindings using \texttt{Pybind11}. 
\subsection{Data Generation}
\label{sec:data-generation}
Because CDRs are data protected by US laws, they are unobtainable. We are unaware of work that models contemporary PSTN call records so we develop a PSTN model to algorithmically generate data. We first generate a graph to represent the \carriers's peering relationships. We then model a social graph of telephone users and assign users to carriers. We then generate CDRs for calls between users. While we believe our model is reasonably accurate, in Section \ref{sec:evaluation}, we show that even if call volumes are significantly higher, \sysname will be practical. 

\myparagraph{Telephone Network} We use an iterative graph generation algorithm to construct a network consistent with real-world telephone topology. We have identified three properties that a reasonable model generator must consider:

\smallspace\noindent\underline{Preferential Attachment}: New carriers prioritize connecting with larger carriers that handle a significant traffic volume, so providers with wider coverage generally acquire more new customers. In our model, the number of carrier connections is proportional to its current degree.

\smallspace\noindent\underline{Market fitness}: Smaller providers can attract new customers but rarely surpass larger providers' market share. This feature enables us to mirror real-world scenarios, such as AT\&T maintaining a higher market share even as the network evolves.

\smallspace\noindent\underline{Inter-carrier agreements}: Represents financial agreements, such as mutual compensation for handling each other's traffic, rates for different types of traffic, and billing arrangements. 

We use the Bianconi-Barabasi model \cite{bianconi2001competition} to achieve these properties. Our network consists of $N$ nodes, each labeled $P_i$ representing a unique carrier node. The weight of an edge between any two carrier nodes signifies the inter-carrier agreement amount, which we use in the shortest path computation. We assume each carrier node seeks to minimize the cost of transmitting call connections, a notion that aligns with real-world practices.

\myparagraph{Subscribers Network} We model subscribers' social interactions using a scale-free network. We create a total of $S$ subscribers, each represented by a phone number in the NPA-NXX-XXXX format. We allocate phone numbers to subscribers based on each carrier's market share. We constructed a Barabási-Albert \cite{barabasi1999emergence} graph denoted as $G_{s}=(V_s, E_s)$ for the subscribers. Since \projectname is primarily interested in scenarios where the caller and the called party belong to different carrier networks, we minimized the probability that neighboring nodes of a given subscriber $s_i$ are on the same network as $s_i$.

\myparagraph{CDR generation} Each edge in the social network $G_{s}$ represents a call between two subscribers $s_{i}$ and $s_{j}$. 
For each call $s_{i}s_{j}$, we represent the call path as the shortest path between their respective providers in the topology. Each hop within the call path represents a CDR record.   \section{Evaluation}
\label{sec:evaluation}
\def\lmsetupmean{0.004}
\def\lmsetupmin{0.003}
\def\lmsetupmax{0.282}
\def\lmsetupstd{0.009}

\def\gmsetupmean{3.152}
\def\gmsetupmin{2.887}
\def\gmsetupmax{4.541}
\def\gmsetupstd{0.140}

\def\tasetupmean{0.113}
\def\tasetupmin{0.102}
\def\tasetupmax{0.196}
\def\tasetupstd{0.010}

\def\carregmean{1.473}
\def\carregmin{1.341}
\def\carregmax{2.055}
\def\carregstd{0.072}

\def\callrate{10000}
\def\Rc{\num[group-separator={,}]{\callrate}}

\def\recrate{10000}
\def\Rrec{\num[group-separator={,}]{\recrate}}

\def\prfevalmean{0.073}
\def\prfevalmin{0.066}
\def\prfevalmax{0.166}
\def\prfevalstd{0.007}
\pgfmathparse{int(1000/\prfevalmean)}
\edef\tempresult{\pgfmathresult}
\def\prfevalspersec{\num[group-separator={,}]{\tempresult}}

\def\grpsiggenmean{1.693}
\def\grpsiggenmin{1.477}
\def\grpsiggenmax{6.573}
\def\grpsiggenstd{0.323}

\def\encmean{2.479}
\def\encmin{2.171}
\def\encmax{3.904}
\def\encstd{0.143}

\def\decmean{0.847}
\def\decmin{0.780}
\def\decmax{1.064}
\def\decstd{0.039}

\def\contrmean{4.143}
\def\contrmin{3.708}
\def\contrmax{5.980}
\def\contrstd{0.173}
\pgfmathparse{int(1000/\contrmean)}
\edef\tempcontr{\pgfmathresult}
\def\contrpersec{\num[group-separator={,}]{\tempcontr}}

\def\tasignmean{0.419}
\def\tasignmin{0.376}
\def\tasignmax{0.615}
\def\tasignstd{0.023}

\def\tracemean{0.419}
\def\tracemin{0.376}
\def\tracemax{0.615}
\def\tracestd{0.023}

\def\gsverifmean{2.310}
\def\gsverifmin{2.118}
\def\gsverifmax{2.865}
\def\gsverifstd{0.098}

\pgfmathparse{int(1000/\gsverifmean)}
\edef\tempverif{\pgfmathresult}
\def\gsverifpersec{\num[group-separator={,}]{\tempverif}}

\def\gsopenmean{0.147}
\def\gsopenmin{0.134}
\def\gsopenmax{0.193}
\def\gsopenstd{0.009}
 For \projectname{} to succeed, runtime performance, queries, and insertions of records need to be fast to handle the volume of call traffic being processed daily by the network. Our prototype implementation allows us to test its performance for each protocol phase. We consider the following metrics: storage growth rate, minimum vCPUs required, time for each protocol, and the minimum bandwidth required as shown in Table \ref{table:sys-req}.

\def\metricscolor{gray}

\footnotesize
\begin{table}[t]
\centering
\caption{Minimum system requirements for each component in \projectname suppose the network processes 10,000 calls per second.}
\label{table:sys-req}
\begin{tabular}{|l|l|l|l|}
\hline
\textbf{Component} & \textbf{Storage} & \textbf{vCPU} & \textbf{Bandwidth} \\ \hline
\multicolumn{4}{|l|}{\textcolor{\metricscolor}{Traceback Authority}} \\\hline
Label Generation & $O(\text{\# of Providers})$ & 4 & 25 Mbps \\
Group Management & $O(\text{\# of Providers})$ & 1 & 20 Mbps \\
Trace Authorization & $O(\text{\# of Traces})$ & 1 & 10 Mbps \\\hline
\multicolumn{4}{|l|}{\textcolor{\metricscolor}{Record Store}} \\\hline
Process Submissions & $O(\text{\# of Records})$ & 257 & 800 Mbps \\\hline
\multicolumn{4}{|l|}{\textcolor{\metricscolor}{All Carriers}} \\\hline
Contribution & $\mathsf{O(1)}$ & 208 & 800 Mbps \\ \hline
\end{tabular}
\end{table}
\normalsize

\myparagraph{Experiment Setup}
Our Experiments were run on a Linux virtual machine with 32 vCPU and 64GB of memory. The host was a Super Micro Server with an Intel Xeon Gold 6130, ECC DDR RAM, and 12Gbps SAS drives. In experiment 1, we benchmark individual tasks such as label generation, record encryption and decryption, opening signatures, signing, and verifying signatures. We executed each process in a single thread 1,000 times to obtain the average, minimum, maximum, and standard deviation (SD) of the runtime.

In Experiment 2, we generated a network graph with 7,000\cite{robomitidb} carriers and simulated $R_C = 10,000$ calls per second. No entity currently has visibility into the call volumes of all carriers in the United States. As a result, we did not find well-supported statistics on the overall call volume in the United States. We are especially concerned with the number of calls that transit multiple carriers, and, of course, this figure is even less attested. Of the statistics we found, many had no citations, did not describe their methodology, or were otherwise suspect in accuracy. As a result, we settled on the round number of 10,000 calls per second for the North American phone network. This corresponds to roughly 800,000,000 calls per day. We admit that this choice is arbitrary, but as we see later, our system has substantial headroom and is also horizontally scalable. Our evaluation in the following sections considers the \TA and \RS as singletons. 

\subsection{Protocol Evaluation}
\myparagraph{\protosetup Protocol} The \protosetup{} is less than 10ms for all entities. Storing the identity of group members (providers) at the \TA grows linearly in the number of providers. 

\footnotesize
\begin{table}[t]
\centering
\caption{Performance for protocols measured in milliseconds.}
\label{table:rec-perf}
\begin{tabular}{|l|r|r|r|r|}
\hline
\bf Task & \bf Mean & \bf Min & \bf Max & \bf Std \\ \hline
Label Generation & \prfevalmean & \prfevalmin & \prfevalmax & \prfevalstd \\ 
Contribution & \contrmean & \contrmin & \contrmax & \contrstd \\ 
Authorization & \tasignmean & \tasignmin & \tasignmax & \tasignstd \\ 
Decryption & 0.847 & 0.780 & 1.064 & 0.039 \\ 
Open & \gsopenmean & \gsopenmin & \gsverifmax & \gsopenstd \\ 
Verify Group Signature & \gsverifmean & \gsverifmin & \gsverifmax & \gsverifstd \\ \hline
\end{tabular}
\end{table}
\normalsize
\myparagraph{\protocontribute{} Protocol}
We measure the time and minimum system requirements to complete the protocol for the label generation and submission phases.

\smallspace\noindent\underline{\textit{Label generation}}:\label{subsec:label-gen}
\Carriers{} request a PRF evaluation from the \TA. Table \ref{table:rec-perf} shows that a single label generation takes \prfevalmean ms on average (SD = \prfevalstd ms). {\em Hence, the \TA can evaluate \prfevalspersec{} labels on average on a single vCPU per second.}

\smallspace\noindent\underline{\textit{Bandwidth for label generation}}: We estimate the minimum bandwidth required to generate labels between \carriers{} and the \TA over an HTTP connection from: 
\begin{equation}\label{eq:1}
B_{w} = R_{rec} \cdot (S_{req} + S_{res}) \cdot (1 + O_{http})
\end{equation} 

\noindent $R_{rec}$ is the rate at which records are generated across the network, $S_{req}$ and $ S_{res}$ are request size, response size, and $O_{http}$ additional overhead (in percentage) introduced by HTTP, respectively. On average, calls generated by our network have 5 hops\footnote{ITG reports that tracebacks usually go through 4 or more hops\cite{itg}}, thus $R_{rec} = 5 \cdot R_C$. Each request payload is 32 bytes; thus, $S_{req} = 256$ bits, likewise, $S_{res} = S_{req}$ since the PRF is length-preserving. We compute overhead using: \begin{equation}
    O_{http} = \text{Overhead per Request} /  \text{Batch Size}
\end{equation}

\noindent We measured the average overhead of about 652 bytes per request for the minimal headers when using HTTP/1.1. The group signature forms a significant fraction of this overhead. {\em The \TA can handle label-generation requests even with substantial network overhead since 25 Mbps is vastly below nominal internet throughput.}

\smallspace\noindent\underline{\it Record Submission}:
We measure the time it takes to contribute 1 CDR record. From Table \ref{table:rec-perf}, contributing 1 CDR takes \contrmean ms on average with a SD of \contrstd ms. A provider can process \contrpersec{} records in a second on a single vCPU. \emph{At a rate of 50,000 records per second\footnote{For an average of 5 hops per call, $10,000$ calls per second correspond to $50,000$ CDRs per second}, a minimum of 208 vCPUs are required to encrypt all records}. This may seem high, but recall we estimated 7,000 providers, so the CPU requirements are small in practice for \carriers.

\smallspace\noindent\underline{\it Bandwidth for record submission}: The record submission request payload is $1,900$ bytes in size ($15,200$ bits), comprising a label, ciphertext, and signature. Using Equation \ref{eq:1}, we estimate the minimum bandwidth for submitting records as 800 Mbps. {\em Therefore, with 800 Mbps the \RS{} can handle submission requests for the entire network.}

Once the \RS{} receives the submission request, it verifies the signature and inserts it into the database. Table \ref{table:rec-perf} shows that verifying a group signature takes \gsverifmean ms on average with a SD of \gsverifstd ms. \emph{Hence, the \RS{} can verify \gsverifpersec{} submissions per core per second, thus requiring a minimum of 24 vCPUs to validate contribution requests}.

\begin{figure}[t]
  \centering
  \includegraphics[width=3in]{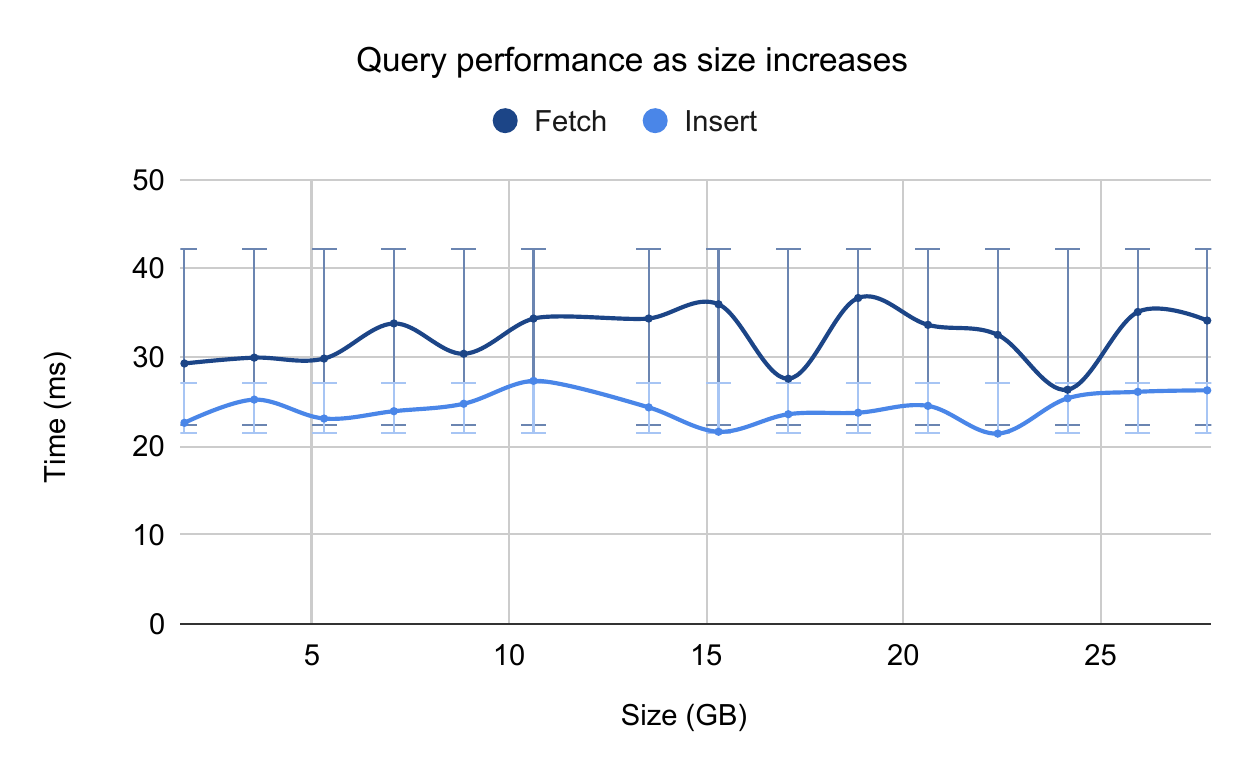}
  \caption{Performance measured in milliseconds for insertion and select queries as database size grows.}
  \label{fig:query-perfs}
\end{figure}

\myparagraph{Record Store}
Our second experiment evaluates the growth of storage and how that affects querying and inserting records. As shown in Figure \ref{fig:query-perfs}, we observed that the time it takes to insert records into the database is independent of the database size, averaging about 24.28ms with a SD of 1.681ms. Note that 23.28ms represents the average time to process a single INSERT statement with 1 VALUE row. Insertion overhead could significantly improve if multiple values are appended to a single insert statement.  {\em The \RS can process 43 insert statements in a second on a single vCPU. To cover the rate at \Rc{} calls per second requires a minimum of 233 vCPUs}. The database size grows at a rate of 1.5 TB per day, roughly \$100 per day. Deployment consideration may require that records are only retained for a given period, after which they get expunged.

\myparagraph{\prototrace{} Protocol}
We consider the \prototrace{} protocol for a single traceback.
Label generation is the same as before.  Whenever authorization for a traceback is requested, the \TA{} records this in the database along with the \carrier{}. Storage grows linearly with the number of traceback requests. To decrypt a record, we need a signature on the label. Signing a label takes a mean time of $0.419$ms with an SD of $0.023$ms. We measured that decrypting a single record takes $0.847$ms on average with a SD of $0.039$ms. \emph{A single vCPU can decrypt 1,180 ciphertexts per second.}

\myparagraph{\protoopen{} Protocol}
In the \protoopen{} protocol, we determine the faulty hops from the decrypted records as described in section \ref{sec:traceback-validation}. We measured the runtime for analyzing the records and determining the faulty hops to be 0.052ms. Opening a signature takes \gsopenmean ms on average with a SD of \gsopenstd ms as shown in Table \ref{table:rec-perf}.

\subsection{Traceback Evaluation}\label{sec:traceback}
\negvspace\myparagraph{Throughput} We estimate the compute time for one traceback as the sum of the following components: 1) label generation, 2) trace authorization, 3) retrieving records from \RS, and 4) decrypting records. In Section \ref{proto:trace}, we generate all labels in the range $\LLL = \{\ttoep(\ts^*) \mid \forall \ts* \in [\ts - t_{\text{max}}, \ts + t_{\text{max}}]\}$. Assuming $t_{\text{max}} = 10$ seconds and $ep$ is in seconds, we retrieve records for 21 labels ($2\cdot t_{max} + 1$). The estimated average compute time for one traceback is 0.75 seconds. \emph{Without communication latency, we can complete close to 3.5 million tracebacks in a month, increasing the current throughput by a multiplicative factor of 11,520}. Communication latency would reduce this number, but the result remains a significant fraction of the current abuse.

\myparagraph{Partial Deployment}
\sysname was explicitly designed to support partial deployment because of lessons learned from S/S and proposals like RPKI to secure Internet communications. Given that both of those systems failed to meet their goals at the current levels of deployment, it is worth considering how \sysname might perform. 

First, though, we will need to establish definitions and deployment models. We say that a traceback is successful if the \sysname record store has at least one record that identifies the originating hop in the call path.
For this analysis, we call providers that deploy \sysname ``adopters,'' and define ``adoption rate'' as the fraction of all providers who are adopters. 
In S/S, the largest networks tended to be the earliest adopters, and only small networks continue operating without S/S (excluding, of course, legacy networks that are incompatible).
This trend held because regulatory agencies gave smaller networks more time to comply with deployment mandates than larger ones.
Published tracebacks and successful enforcement actions report that the vast majority of illegal calls come from small networks. 

We used our network model from Section~\ref{sec:data-generation} to estimate traceback success in partial deployment. We simulate randomly dialed robocalls originating from the smallest $r\%$ of carriers and adopters as the largest $a\%$ of carriers. We find that if robocalls originate from the bottom 10\% of networks, with \sysname adoption by only the largest 2\% of carriers, \sysname can still successfully traceback \emph{27\%} of all robocalls. When adoption increases to 10\%, which is still lower than the current rate of adoption of S/S, traceback success leaps to 55\% of robocalls. 

Given that robocallers place millions or billions of calls, even at low adoption \sysname would produce mountains of evidence that could lead to takedown of illegal robocalling operations. Of course, these results are likely sensitive to our model choices. Still, even if this estimate is wrong by an order of magnitude, tracing even a small fraction of the thousands or millions of robocalls is a vast improvement compared to 300 per month we see today.

 \section{Deployment Considerations}
\label{sec:discussion}
In this section, we discuss practical deployment concerns relating to incentives, engineering concerns, and the trace authorization process.

\myparagraph{Deployment Incentives} When integrating a new security mandate into an existing system, it is essential to consider incentives. How will a provider benefit from deploying \sysname? Eliminating bulk illegal calls aligns with providers’ interests since these calls are often unanswered and do not generate revenue. However, this motivation alone may not be enough. Fortunately, the S/S framework has shown that regulatory mandates can drive widespread network changes.

\myparagraph{Engineering Concerns} In our discussions with practitioners, we identified engineering concerns that, while challenging, are manageable. We anticipate that carriers will face hurdles integrating with billing systems and managing new cryptographic infrastructure. However, we anticipate that deployment will still be easier than S/S because it involves only latency-insensitive backends. Jäger itself will face typical site reliability engineering challenges, such as replication and public key infrastructure issues like governance. The industry has successfully navigated similar challenges with S/S, so we can have confidence that these engineering concerns are surmountable.

\myparagraph{Rate Limiting and Authorization} Our performance evaluation assumed that traceback authorization would be instantaneous. In reality, there will be some processing time. Human intervention may be required to review requests 
before or after they are system-signed. In some cases, the trace authorizer could
automatically approve requests, such as requests from honeypots for calls to their own numbers. Given that honeypots receive thousands of calls daily \cite{usenix_snorcall, usenix_whoscalling}, scalable traceback will significantly enhance investigative processes. \section{Related Work}
\label{sec:related-work}

Telecom fraud~\cite{sahinsoktele} is a long-standing problem~\cite{mavmobiattack, mcinnes2019analysis, mustafa2014you} that continues to impact carriers~\cite{sahin2016over,reaves2015boxed,murynets2014simbox, sahin2021understandingIRSF} and subscribers~\cite{raviwangiri, arafatwangiri}.
Most fraud schemes stem from inadequate authentication mechanisms~\cite{reaves2016authloop, choi2005improvement, dacosta2010proxychain, reaves2017authenticall, tu2016toward} and security flaws in legacy telecom signaling protocols~\cite{ullahss7, jensen2016better, welch2017exploiting, rehman2014security}.
Attempts to address these problems through protocol enhancements~\cite{sahinsoktele,tu2016sok, gupta2015phoneypot} and defenses~\cite{reaves2015boxed, ohpreventing} have had limited success. Illegal robocalling, a form of telecom fraud, frustrates phone users, carriers, and regulators.
Over the past decade, widespread adoption of VoIP technology has led to a surge of scams~\cite{tu2016sok, sherman2020you,usenix_whoscalling, usenix_snorcall} perpetrated using robocalls.
To study the operational characteristics of robocallers, researchers have employed a wide range of techniques such as CDR data mining~\cite{tseng2015fraudetector, liu2018augmenting, 8804189}, machine learning~\cite{5270323, li2018machine, gowri2021detection}, audio processing~\cite{usenix_whoscalling, usenix_snorcall}, carrier collaboration~\cite{azad2017privy, azad2019rapid}, and reputation scoring ~\cite{bokharaei2011you, 4595872}. 
Mitigation techniques based on caller ID authentication~\cite{tu2017toward}, spam filtering~\cite{dantu2005detecting}, call-blocking apps~\cite{mustafa2014you, robohalt, Truecaller, Hiya, Aura_Call_Assistant}, and increased penalties~\cite{fcc_tcpa_rules,tracedact} have been proposed.
However, they have failed to significantly deter bad actors from originating illegal robocalls. 

In Dec 2019, the US Congress passed the TRACED Act~\cite{tracedact} to protect consumers from illegal robocalls. Consequently, the FCC designated the Industry Traceback Group (ITG)~\cite{itg} to track down entities responsible for originating illegal robocall traffic using traceback.
Tracebacks remain invaluable in uncovering and prosecuting numerous illegal robocalling operations~\cite{fcc_investigation_1}.
However, its effectiveness is limited since it is a manual, iterative, and time-consuming process. Each traceback requires cooperation among carriers spanning multiple days to pinpoint the source of illegal robocall traffic.
Network traceback methods like packet marking and router logging~\cite{4447464, 1195496, 1494507} are ineffective to traceback phone calls~\cite{hsu2011collaborative} due to general IP traceback limitations.

Our automated traceback technique addresses these challenges and encompasses all transit carriers. 
Notably, \projectname does not require modifications to existing infrastructure, making it compatible with other protocols. \section{Conclusion}
\label{sec:conclusion}

In this paper, we described the design of \sysname, a distributed system to facilitate automatic call traceback. \sysname facilitates the anonymous-but-traceable submission of encrypted call records to a central source, which after vetting from an authorizer allows traceback only by parties with information about a call to be traced. We demonstrate that despite the expensive cryptographic primitives and coordination cost, the system is practical today with modest hardware and low latency. 
In so doing, we show that \sysname represents a powerful new tool to combat telephone abuse.

 \begin{acks}
We thank our anonymous shepherd and reviewers for their support of the paper. 
This material is based upon work supported by the National Science Foundation under Award No. CNS-2142930. Funds from the 2020 Internet Defense Prize also supported portions of this work.
\end{acks} 
\bibliographystyle{abbrv}
\balance
\bibliography{papers}

\ifisFullVersion
    \appendix

\begin{figure*}[!htbp]

\begin{tcolorbox}[
    standard jigsaw,
    opacityback=0]{
    
    \underline{\bf Functionality {$\fprivtb$}}
    \vspace{2pt}

    \textbf{Entities:}  $\PPP = \emptyset$ is the set of carriers, $\MMM$ is the set of malicious carriers, $\TA$ is a traceback authority
    
    This functionality is initialized with predicate $\validate$ which determines if a list of records has conflicts. It takes as inputs a list of $\cdr = (P_{i-1}\|P_i\|P_{i+1})$ and outputs a set conflicting $\cdr$ (denoted $\CCC_\conflicts$)

    \textbf{Data Structures}: $\DDD$ is the table of records, that stores entries of the form $(\ctime: P_i, \callpp, \cdr)$ 

    \vspace{4pt}

    \underline{\textbf{Interface}}:

    \begin{itemize}

    \item {\bf Register}: Upon receiving $(\cregister,P_i)$ from a party $P_i$, do $\PPP = \PPP \cup \{P_i\}$ and send $(\cregister, P_i)$ to $\adv$. 

    \item  {\bf Contribute Call}: Upon receiving $(\ccall, \callpp, \cdr)$ from some carrier $P_i$, check if $P_i \in \PPP$. If yes,  store $(\ctime: P_i, \callpp, \cdr)$ in $\DDD$. Send $(\ccall, \ctime)$ to the $\adv$. Upon receiving $(\ccall, \ok)$ from $\adv$, send $(\ccall, \ctime)$ to $P_i$. Update $\ctime = \ctime + 1$.

    \item {\bf Malicious Update to Database}: Upon receiving $(\malupdate,(\delete,\ctime))$ from $\adv$, update $\DDD \setminus \{(\ctime: (\cdot))\}$. Upon receiving $(\malupdate,(\add,(\cdot): (P_j, \callpp, \cdr)))$ from $\adv$,  update $\DDD \cup \{(\ctime: (P_j, \callpp, \cdr))\}$ and send $\ctime$ to $\adv$. 
    
   \item  {\bf Trace Call}: Upon receiving $(\ctrace, \callpp, P_i)$ from $P_i$,
   \begin{enumerate}
       \item  Retrieve all entries of the form  ($\ctime: \callpp^*, \cdot, \cdot$) in $\DDD$ where the $\ts^*$ of $\callpp^*$ belongs to [$\ts-max, \ts+max$], where $max$ is the max size for a call and $\ts$ is the timestamp in $\callpp$. Let $\CCC_\trace = \{(\ctime, \cdr)\}$ for each retrieved $\cdr$ 
       \item If there are no $\cdr$ missing in $\CCC_\trace$ send the set of retrieved $(\{\ctime\})$ to $\adv$. If there are $\cdr$ missing, then send $(\callpp,\{\ctime\})$ to $\adv$.
       Receive $(\{\malupdate, (\add, \ctime_\adv: P_j,\callpp_\adv, \cdr_\adv)\})$ or $(\malupdate, (\delete, \ctime))$. If $P_j \in \MMM$, add these entries to  $\DDD$, else ignore these messages.  Let $\CCC_\trace = \CCC_\trace \cup \malcdr$, where $\malcdr = \{\ctime_\adv, \cdr_\adv\}$. 
       
       \item Send $(\allowtrace, P_i )$ to $\TA$. Upon receiving $(\allowtrace, P_i, \ok)$ from $\TA$, send $\CCC_\trace$ to $P_i$ else send $\emptyset$ to $P_i$
\end{enumerate}

   \item {\bf Open}: Upon receiving $(\copen, \callpp,\CCC_\trace)$ from some party $P_i$, if $\TA$ is corrupt, send $(\copen, \callpp,\CCC_\trace)$ to $\adv$, and receive $(\copen, \{P^*\})$. Else: 
   \begin{enumerate}
       \item Compute $\CCC_\conflicts = \validate(\CCC_\trace)$
       \item For each $\cdr^* \in \CCC_\conflicts$ retrieve $P*$ such that ($P^*,\callpp, \cdr, \cdot$) in $\DDD$
       
   \end{enumerate}

 Return $(\copen, \{P^*\})$  to the calling $P_i$
    \end{itemize}
 }

\end{tcolorbox}
\caption{Private Traceback functionality}
    \label{fig:ptb}
\end{figure*}

\section{Ideal functionality for \projectname}
In this section we formalize the security properties of \projectname~ in the UC framework \cite{canetti2001new}. We define an ideal functionality $\fprivtb$ (Figure~\ref{fig:ptb}) that captures the correctness and the security properties of the \projectname~ system. 

The $\fprivtb$ functionality maintains a database $\DDD$ 
and provides the following interface: 
\begin{itemize}
    \item $\cregister$: Enables carriers to register with the system. Since this is public information, the identity of the carrier is leaked to the adversary. 
    \item $\ccall$: Allows carriers to submit records to the system. Recall that in the real world, an adversary can always learn when a record is submitted but does not learn the contents of the record, nor the identity of the carrier that submits the record. Therefore, the only information that is leaked to the adversary is the $\ctime$ which gives an indication of what time a call record was submitted. This captures the \textit{anonymity} and the \textit{confidentiality} guarantees. 
    \item $\malupdate$: Allows the adversary to delete or add records to the database. 
    \item $\ctrace$: Enables carriers to retrieve the $\cdr$s relevant to a specific call. All the $\cdr$s that are currently in the database along with any $\cdr$ that the adversary wants to append are returned to the carrier. Before sending these $\cdr$s to the carrier, the functionality sends $\allowtrace$ command to the $\TA$, and only if the $\TA$ responds with $(\allowtrace, \ok)$ are these $\cdr$s sent to the carrier. This captures the \textit{trace authorization} requirement and ensures that a carrier cannot request a trace too many times and hence \textit{rate-limits} their requests. 
    \item $\copen$: Allows a carrier to deanonymize the sender of $\cdr$s that are malformed or conflicting. If the $\TA$ is honest, the functionality runs a $\validate$ predicate to determine the malicious/conflicting hops and returns the identities of the corresponding carriers. On the other hand if the $\TA$ is malicious, the adversary is allowed to return the identities of any of the  carriers. This captures the property that \textit{accountability} is guaranteed as long as the $\TA$ is honest.  
\end{itemize}

\section{The \projectname protocol}\label{prot:details}
We need the following ingredients: 
\begin{enumerate}
    \item Group signatures as defined in \cite{bootle2016foundations} 
    \item An OPRF scheme
    \item Witness Encryption scheme for signatures. 
\end{enumerate}

In the presentation of \projectname~ below we will instantiate the OPRF scheme using the 2HashDH OPRF scheme of Jarecki et. al. \cite{jarecki2017toppss} and the witness encryption scheme presented in \cite{wes}\cite{mcfly}. 

\projectname~ consists of four protocols:  \emph{\protosetup}, \emph{\protocontribute}, \emph{\prototrace}, and \emph{\protoopen} protocols, which we describe in detail below. 

Before we describe the protocol we present details of the OPRF scheme of \cite{jarecki2017toppss} in Figure~\ref{fig:oprf-scheme} and the Witness Encryption scheme of \cite{wes} in Figure~\ref{fig:wes}.

\begin{figure}
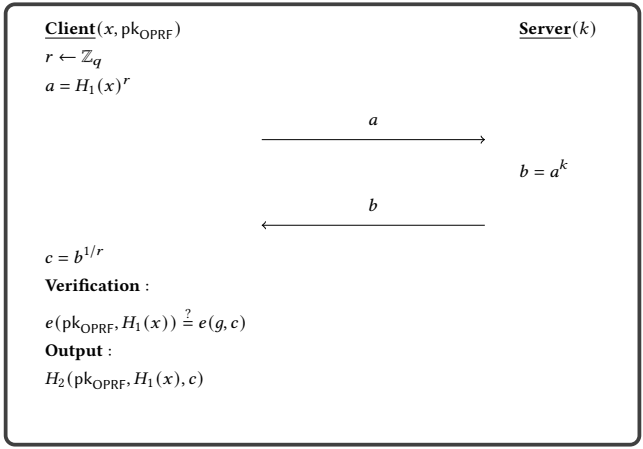

    \centering
    \begin{tcolorbox}[
    standard jigsaw,
    opacityback=0]
\begin{adjustbox}{max width=\textwidth,max totalheight=\textheight, center}
    \pseudocode{\textbf{\underline{Client}}(x,\pk_\oprf) \>\> \textbf{\underline{Server}}(k)\\
            r \gets \ZZ_q \\
            a = H_1(x)^r\\
            \> \sendmessageright*{a} \> \\
            \>\> b = a^k \\
            \> \sendmessageleft*{b}\> \\
            c = b^{1/r} \\
            \textbf{Verification}: \\
            e(\pk_\oprf,H_1(x)) \stackrel{?}{=} e(g, c) \\
            \textbf{Output}: \\
             H_2(\pk_\oprf, H_1(x), c) \\
            }
        \end{adjustbox}
    \end{tcolorbox}
    \caption{OPRF Scheme of \cite{jarecki2017toppss} where $\pk_\oprf = g^k$ and is previously announced by the server. All groups are pairing-friendly. }
    \label{fig:oprf-scheme}
\end{figure}

\begin{figure}
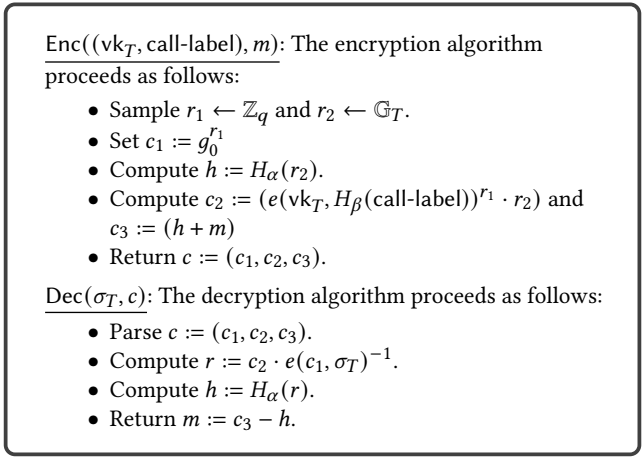

    \centering
    \begin{tcolorbox}[
    standard jigsaw,
    opacityback=0]

    \begin{flushleft}
    
    \smallskip\noindent\underline{$\enc((\vk_T, \clabel), m)$}: The encryption algorithm proceeds as follows:
    \begin{itemize}
      \item Sample $r_1 \gets \ZZ_q$ and $r_2 \gets \GG_T$.
        \item Set $c_1 := g_0^{r_1}$
        \item Compute $h := H_\alpha(r_2)$. 
        \item Compute $c_2 := (e(\vk_T, H_\beta(\clabel))^{r_1}\cdot r_2)$ and $c_3 := (h + m)$
        \item Return $c := (c_1, c_2, c_3)$.
    \end{itemize}
     
     \smallskip\noindent\underline{$\dec(\sigma_T, c) $}: The decryption algorithm proceeds as follows:
     \begin{itemize}
         \item Parse $c := (c_1, c_2, c_3)$.
         \item Compute $r := c_2 \cdot e(c_1, \sigma_T)^{-1}$.
         \item Compute $h := H_\alpha(r)$.
         \item Return $m := c_3 - h$.
     \end{itemize}
    \end{flushleft}
    \end{tcolorbox}
    \caption{Witness encryption based on BLS signatures from \cite{wes}, here $H_\alpha$ and $H_\beta$ are hash functions modeled as random oracles. }
    \label{fig:wes}
\end{figure}

\subsection{\protosetup} 

The Traceback Authority $\TA$ does: 
\begin{enumerate}
    \item Generate BLS signature keys $(\sk_T, \vk_T) \gets \sign.\kgen(1^\lambda)$ and $(\sk_R, \vk_R) \gets \sign.\kgen(1^\lambda)$
    \item Generate OPRF key $k \gets \ZZ_q$ and announce the corresponding public key $\pk_\oprf = g^k$
    \item Run the $\gkgen$ algorithm of the group signature scheme and announce $(\gpk, \info_0)$  which are the group manager public key and  the initial group information. 
    \item The $\TA$ initializes a counter $\rlctr_i$ corresponding to each $P_i$. 
\end{enumerate}

Each provider $P_i$ does: 
\begin{enumerate}
    \item Run the interactive joining protocol $\gjoin$ with $\TA$ and receive $\gsk_i$ 
    \item Generate signing keys $(\sk_i, \vk_i) \gets \sign.\kgen(1^\lambda)$ and announce $\vk_i$. 
\end{enumerate}

The Record Store $\RS$ does: 
\begin{enumerate}
    \item Initialize the database $\DDD$
    \item Generate signing keys $(\sk_{\RS}, \vk_{\RS}) \gets \sign.\kgen(1^\lambda)$ and announce $\vk_{\RS}$
\end{enumerate}

\subsection{\protocontribute}

Each provider $P_i$ with input $(\src\|\dst\|\ts)$ and $\cdr = (P_{i-1}\|P_i\|P_{i+1})$ does: 
\begin{enumerate}
    \item Compute $\ep = \ttoep(\ts)$ and run the OPRF protocol with $\TA$ using inputs  $\callpp = (\src\|\dst\|\ep)$ as follows: 
    \begin{enumerate}
        \item Pick $r\gets\ZZ_q$ and compute $a = H_1(\callpp)^r$ and send $a$ to $\TA$
        \item The $\TA$ computes $b = a^k$ and sends it back to $P_i$
        \item $P_i$ checks the following pairing equation to verify the OPRF was evaluated correctly: $e(\pk_\oprf, H_1(\callpp)) = e(g, b^{1/r})$\footnote{The groups are pairing friendly, and hence we can verify the correctness of the OPRF via pairing equations \cite{jarecki2017toppss}}.
\item Output $\clabel = H_2(\pk_\oprf, \callpp, b^{1/r} )$ 
    \end{enumerate}
    \item The provider then encrypts the $\cdr$ as follows:
    \begin{enumerate}
        \item Sample a random key $\cdrkey \gets \{0,1\}^\lambda$
        \item Encrypt $\cdrkey$ as as $\ct_1 = \WE.\enc((\vk_T, \clabel), \cdrkey)$
        \item Compute $\ct_2 = H_3(\callpp\|\cdrkey) \oplus \cdr$ 
    \end{enumerate}
    \item The provider signs the ciphertexts and the $H(\clabel)$ using the group signature scheme: $\sigma = \gsign(\gsk_i, (\ct_1, \ct_2, H(\clabel)))$ and sends $(H(\clabel), (\ct_1, \ct), \sigma)$ to $\RS$
\end{enumerate}

The Record Store $\RS$ upon receiving  $(H(\clabel), (\ct_1, \ct), \sigma)$ does the following:
\begin{enumerate}
    \item Check $\gverify(\gpk, (H(\clabel), (\ct_1, \ct)), \sigma) = 1$. 
    \item If yes, write $(H(\clabel), (\ct_1, \ct), \sigma)$ to the database. Else ignore the message. 
\end{enumerate}

\subsection{\prototrace}

Each provider $P_i$ with input $(\src\|\dst\|\ts)$ does: 
\begin{enumerate}
    \item For $\ts* \in [\ts - \tmax, \ts + \tmax]$ compute $\ep^* = \ttoep(\ts^*)$ 
    \item For each $\ep*$ compute $\clabel^*$ as above with inputs $\callpp^*\allowbreak = (\src\|\dst\|\ep^*)$ by running the OPRF protocol with the $\TA$
    \item Request a signature on $\idx = H(\clabel^*)$ from $\TA$. 
    \item The $\TA$ checks if $\rlctr_i > T$, if yes, reject the request, else the $\TA$ computes $\sigma_R = \sign(\sk_R, \idx)$ and sends it back to $P_i$. 
    \item Send $(\idx^*, \sigma_R)$ to $\RS$ and receive the corresponding records back $((\idx^*, \ct_1^*, \ct_2^*, \sigma^*), \sigma_{\RS})$ if they exist.
    \item Verify $\verify(\vk_{\RS},((\idx^*, \ct_1^*, \ct_2^*, \sigma^*), \sigma_{\RS})) = 1$. If not, reject. 
    \item Request a signature on each $\clabel^*$ from $\TA$ by sending $\sigma_i = \sign(\sk_i, \clabel^*)$ to the $\TA$. The $\TA$ compute $\sigma_T = \sign(\sk_T, \clabel^*)$ and sends it to $P_i$.

    \item Decrypt the ciphertexts as follows:
    \begin{enumerate}
        \item Compute $\cdrkey^* = \WE.\dec(\sigma_T, \ct_1^*)$
        \item Compute $\cdr^* = H_3(\callpp^*\|\cdrkey^*) \oplus \ct_2$
    \end{enumerate}
    \item Append $\cdr^*$ to $\CCC_\trace$
\end{enumerate}

\subsection{\protoopen}

If a $\cdr^*$ is malformed or conflicts with another record, the provider $P_i$ can request the $\TA$ to open the corresponding signature to deanonymize that provider and hold them accountable. 

\begin{enumerate}
    \item $P_i$ sends $\callpp,\clabel^*, \ct_1^*, \ct_2^*, \sigma^*)$, the list of $\{\cdr\}$  to the $\TA$
    \item The $\TA$ runs $\validate$ algorithm  to determine if the provider is misbehaving and needs to be deanonymized. 
    \item The $\TA$ computes $P_j^* = \gopen(\gsk, (\clabel^*, \ct_1^*, \ct_2^*, \sigma^*))$ and returns $P_j^*$
\end{enumerate}

\section{Formal Proofs of Security of \projectname~}\label{sec:ucproof}

In this section we will present the formal proofs of security. We will consider three cases of corruption: (1) Only a subset of the providers are corrupt (2) The record store $\RS$ is corrupt and can collude with any of the providers and (3) the traceback authority $\TA$ is corrupt and can collude with any of the providers. 

For completeness we first show a simulator where no parties are corrupt. 

\myparagraph{Case 0: No entities are corrupt}

\myparagraph{\protosetup}
Simulate the  Traceback Authority: 
\begin{enumerate}
    \item Generate BLS signature keys $(\sk_T, \vk_T) \gets \sign.\kgen(1^\lambda)$ and $(\sk_R, \vk_R) \gets \sign.\kgen(1^\lambda) $
    \item Generate OPRF key $k \gets \ZZ_q$ and announce the corresponding public key $\pk_\oprf = g^k$
    \item Run the $\gkgen$ algorithm of the group signature scheme and announce $(\gpk, \info_0)$  which are the group manager public key and  the initial group information. 
\end{enumerate}

Simulate honest providers: Upon receiving $(\cregister, P_i)$ from $\fprivtb$: Generate signing keys $(\sk_i, \vk_i) \gets \sign.\kgen(1^\lambda)$ and announce $\vk_i$.

\myparagraph{\protocontribute}

Simulating honest contributions: Upon receiving $(\ccall,\allowbreak \ctime)$ from $\fprivtb$, just store $(\ctime, (\cdot))$ in a database $\DDD$ and return $(\ccall, \ok)$ to $\fprivtb$.

\myparagraph{\prototrace}  

Honest trace request:
\begin{enumerate}
    \item Upon receiving $(\ctrace,\ctime)$ from $\fprivtb$, send $\emptyset$ back to $\fprivtb$
    \item Upon receiving $(\allowtrace, P_i)$ from $\fprivtb$, send $(\allowbreak\allowtrace, P_i, \ok)$ back to $\fprivtb$. 
\end{enumerate}

\subsection{Case 1: Only a subset of the providers are corrupt and the \TA and \RS are honest}
To prove UC security we need to show that there exists a simulator that produces a transcript in the ideal world that is indistinguishable from the real world. Below we present the simulator for the case when only a subset of the providers are malicious and all other entities are honest. 

\myparagraph{\protosetup}
Simulate the  Traceback Authority: 
\begin{enumerate}
    \item Generate BLS signature keys $(\sk_T, \vk_T) \gets \sign.\kgen(1^\lambda)$  and $(\sk_R, \vk_R) \gets \sign.\kgen(1^\lambda) $
    \item Generate OPRF key $k \gets \ZZ_q$ and announce the corresponding public key $\pk_\oprf = g^k$
    \item Run the $\gkgen$ algorithm of the group signature scheme and announce $(\gpk, \info_0)$  which are the group manager public key and  the initial group information. 
\end{enumerate}

Simulate honest providers: Upon receiving $(\cregister, P_i)$ from $\fprivtb$: Generate signing keys $(\sk_i, \vk_i) \gets \sign.\kgen(1^\lambda)$ and announce $\vk_i$. 

Malicious provider $P_j$: Simulate the interactive joining protocol $\gjoin$ with $P_j$, and send $(\cregister, P_i)$ to $\fprivtb$. 

\myparagraph{\protocontribute}

Simulating honest contributions: The simulator does not need to simulate interactions between honest providers and $\RS$ and $\TA$, since this is not in the view of the malicious providers. Therefore, upon receiving $(\ccall, \ctime)$ from $\fprivtb$, just store $(\ctime, (\cdot))$ in a database $\DDD$ and return $(\ccall, \ok)$ to $\fprivtb$.

Simulating random oracle invocations: 
\begin{enumerate}
    \item Upon receiving input $x$ for random oracle $H_1$, check if $(x, y)$ exists in $\QQQ_1$. If yes return $y$, else sample a random $y$, store $(x,y) \in \QQQ_1$ and return $y$. 
    \item Upon receiving input $x$ for random oracle $H_2$, check if $(x, y)$ exists in $\QQQ_2$. If yes return $y$, else sample a random $y$, store $(x,y) \in \QQQ_2$ and return $y$. 
     \item Upon receiving input $x$ for random oracle $H_3$, check if $(x, y)$ exists in $\QQQ_3$. If yes return $y$, else sample a random $y$, store $(x,y) \in \QQQ_3$ and return $y$. 
\end{enumerate}

Simulating malicious contributions: 
\begin{enumerate}
    \item Upon receiving $a$ on behalf of $\TA$ from the adversary, compute $b = a^k$ and send it back to the $\adv$. 
    \item Upon receiving $(\clabel, \ct_1, \ct_2, \sigma)$ from $\adv$:
    \begin{enumerate}
        \item Check $(x, \clabel)$ exists in $Q_2$. If not, abort with $\rofail_2$. 
        \item Else parse $x$ as $(\pk_\oprf, \callpp, b^*)$ 
        \item Check that $(\callpp, y)$ exists in $Q_1$. If not, abort with $\rofail_1$. 
        \item Check that $e(\pk_\oprf, y) = e(g, b^*)$. If not, abort with $\prffail$
        \item Check $((\callpp\|\cdrkey^*), z)$ exists in $\QQQ_3$. If not, abort with $\rofail_3$. 
        \item Else compute $\cdr^* = z \oplus \ct_2$ 
\end{enumerate}
    \item If all checks pass, compute $P_j^* = \gopen(\gsk_i, (\clabel, \ct_1, \ct_2, \sigma))$. If $P_j^*$ corresponds to that of an honest party, abort with $\groupsigfail$. 
    \item Send $(\ccall, \callpp, \cdr)$ on behalf of $P_j^*$ to $\fprivtb$. 
    \item Receive $(\ccall, \ctime)$ from $\fprivtb$ and store $(\ctime, (P_j, \callpp, \cdr))$ in $\DDD$. 
\end{enumerate}

\myparagraph{\prototrace} 
We consider two cases: 1) an honest trace request 2) a malicious trace request 

\begin{enumerate} 
\item Honest trace request:
\begin{enumerate}
    \item Upon receiving $(\ctrace,\ctime)$ from $\fprivtb$, send $\emptyset$ back to $\fprivtb$
    \item Upon receiving $(\allowtrace, P_i)$ from $\fprivtb$, send $(\allowtrace, P_i, \ok)$ back to $\fprivtb$. 
\end{enumerate}

\item Malicious trace request:
\begin{enumerate}
    \item Upon receiving $(\clabel^*, \sigma^*)$ from $\adv$ (on behalf of $P_j^*$): 
    \begin{enumerate}
    \item if $\sigma^*$ verifies under a honest party $P_i$'s $\vk$ abort with $\sigfail$ 
        \item If $\clabel^*, (\cdot)$ exists in the list of ciphertexts, send $(\clabel^*, (\ct_1, \ct_2, \sigma))$ to $\adv$. Else:
        \item Check $(x, \clabel)$ exists in $Q_2$. If not, abort with $\rofail_2$. 
        \item Else parse $x$ as $(\pk_\oprf, \callpp, b^*)$ 
        \item Check that $(\callpp, y)$ exists in $Q_1$. If not, abort with $\rofail_1$. 
        \item Check that $e(\pk_\oprf, y) = e(g, b^*)$. If not, abort with $\prffail$
    \end{enumerate}
    \item Send $(\ctrace, \callpp, P_j^*)$ on behalf of $P_j^*$ to $\fprivtb$ 
    \item Receive $(\{\ctime\})$ from $\fprivtb$. If any of the $\ctime$ correspond to that of a malicious contribution, send $\{\ctime, P_j, \callpp, \cdr\}$ to $\fprivtb$. 
    \item Upon receiving $\allowtrace$ from $\fprivtb$, check that requesting provider has not requested too many records and send $(\allowtrace, \ok)$ to $\fprivtb$. 
    \item Receive $\CCC_\trace$ from $\fprivtb$. 
    \item Encrypt each of the $\cdr$ using the corresponding $\clabel$ and $\vk_T$ in the following way:
    \begin{enumerate}
        \item Sample a random key $\cdrkey \gets \{0,1\}^\lambda$ 
        \item Compute $\ct_1 = \WE.\enc((\clabel, \vk_T), \cdrkey)$
        \item  Sample a random  $\ct_2$. Set $H_3(\callpp\|\cdrkey) = \ct_2\oplus \cdr$
    \end{enumerate}
    \item Using the group signature scheme, compute $\sigma$ on behalf of $P_i$, where $P_i$ is the honest sender of the record. 
    \item Compute $\sigma_{RS} = \sign((\clabel^*, \ct_1^*, \ct_2^*, \sigma^*))$
    \item Send the $(\clabel^*, \ct_1^*, \ct_2^*, \sigma^*), \sigma_{RS}$ that correspond to $\clabel^*$ to the adversary. 
\end{enumerate}

\end{enumerate}

\myparagraph{\protoopen} Upon receiving $\gopen$ request from $\adv$ for a particular record, $(\clabel^*, \ct_1^*, \ct_2^*, \sigma^*), \sigma_{RS}$, check that $\sigma_{RS}$ is a signature that was computed by the simulator, if not abort with $\sigfail_2$. Else send $\copen, \callpp, \CCC_\trace$ to $\fprivtb$ and output whatever $\fprivtb$ returns.

\myparagraph{Proof By Hybrids} Now to prove that the simulated world and the real world are indistinguishable we proceed via a sequence of hybrids, starting from the real world until we reach the ideal world. We show that each of these hybrids are indistinguishable and therefore the real world and the simulated world are indistinguishable. 

\begin{itemize}
    \item[$\hyb_0$] This is the real world protocol 
    \item[$\hyb_1$] This hybrid is identical to the previous hybrid except that the simulator may abort with $\groupsigfail$. By the non-frameability property of the group signature scheme, the simulator aborts with negligible probability and therefore this hybrid is indistinguishable from the previous one. 
    \item[$\hyb_2$] This hybrid is identical to the previous hybrid except that the simulator may abort with $\rofail_1$. Since we use a random oracle and require the adversary to use the RO, the probability of this event occurring is negligible, and therefore this hybrid is indistinguishable from the previous one. 
    \item[$\hyb_3$] This hybrid is identical to the previous hybrid except that the simulator may abort with $\rofail_2$.Since we use a random oracle and require the adversary to use the RO, the probability of this event occurring is negligible, and therefore this hybrid is indistinguishable from the previous one. 
    \item[$\hyb_4$] This hybrid is identical to the previous hybrid except that the simulator may abort with $\rofail_3$. Since we use a random oracle and require the adversary to use the RO, the probability of this event occurring is negligible, and therefore this hybrid is indistinguishable from the previous one. 
    \item[$\hyb_5$] This hybrid is identical to the previous hybrid except that the simulator may abort with $\sigfail$. Since we use unforgeable signatures, this event occurs with negligible probability and therefore the two hybrids are indistinguishable. 
    \item[$\hyb_6$] This hybrid is identical to the previous hybrid except that the simulator may abort with $\sigfail_2$. Since we use unforgeable signatures, this event occurs with negligible probability and therefore the two hybrids are indistinguishable. 
\end{itemize}

Since this hybrid is identical to the simulated world, we have shown that the real world and ideal world are indistinguishable, and that concludes the proof of security for the case when only a subset of carriers are corrupt.

\subsection{Case 2: \TA and subset of providers are corrupt, and \RS is honest}
We present the simulator for this case below: 

\myparagraph{\protosetup}
The adversary $\adv$ runs the algorithms of the $\TA$. Receive $\vk_T, \pk_\oprf, (\gpk, \info_0)$  from $\adv$

Simulate honest providers: Upon receiving $(\cregister, P_i)$ from $\fprivtb$: 
\begin{enumerate}
    \item Generate signing keys $(\sk_i, \vk_i) \gets \sign.\kgen(1^\lambda)$ and announce $\vk_i$. 
    \item Interact with $\adv$ to run the $\gjoin$ protocol and learn $\gsk_i$.  
\end{enumerate}

Malicious provider $P_j$: Upon receiving  $(\updgrp, P_j^*)$ from $\adv$ on behalf of $\TA$, send $(\cregister, P_j^*)$ to $\fprivtb$.

\myparagraph{\protocontribute}

Simulating honest contributions: Note that the simulator only simulates the computation of the label, since that is the only interaction with the adversary. Since the $\RS$ is honest the simulator does not need to compute the ciphertexts or interact with $\adv$. Therefore, upon receiving $(\ccall, \ctime)$ from $\fprivtb$: 
\begin{enumerate}
    \item Compute a label as follows: 
    \begin{enumerate}
        \item Sample random $r \gets \ZZ_q$ and compute $a = H_1(0)^r$ and send $a$ to $\adv$ on behalf of $\TA$. 
        \item Receive $b$ from $\adv$. 
        \item Check that $e(\pk_\oprf, H_1(0)) = e(g, b^{1/r})$     
        \item Output $\clabel = H_2(\pk_\oprf, 0, b^{1/r})$   and store $(\ctime, \clabel)$. 
    \end{enumerate}
\end{enumerate}

Simulating random oracle invocations: 
\begin{enumerate}
    \item Upon receiving input $x$ for random oracle $H_1$, check if $(x, y)$ exists in $\QQQ_1$. If yes return $y$, else sample a random $y$, store $(x,y) \in \QQQ_1$ and return $y$. 
    \item Upon receiving input $x$ for random oracle $H_2$, check if $(x, y)$ exists in $\QQQ_2$. If yes return $y$, else sample a random $y$, store $(x,y) \in \QQQ_2$ and return $y$. 
     \item Upon receiving input $x$ for random oracle $H_3$, check if $(x, y)$ exists in $\QQQ_3$. If yes return $y$, else sample a random $y$, store $(x,y) \in \QQQ_3$ and return $y$. 
\end{enumerate}

Simulating malicious contributions: 
\begin{enumerate} 
    \item Upon receiving $(\clabel, \ct_1, \ct_2, \sigma)$ from $\adv$:
    \begin{enumerate}
        \item Check $(x, \clabel)$ exists in $Q_2$. If not, abort with $\rofail_2$. 
        \item Else parse $x$ as $(\pk_\oprf, \callpp, b^*)$ 
        \item Check that $(\callpp, y)$ exists in $Q_1$. If not, abort with $\rofail_1$. 
        \item Check that $e(\pk_\oprf, y) = e(g, b^*)$. If not, abort with $\prffail$
        \item Check $((\callpp\|\cdrkey^*), z)$ exists in $\QQQ_3$. If not, abort with $\rofail_3$. 
        \item Else compute $\cdr^* = z \oplus \ct_2$ 
    \end{enumerate}
    \item Send $(\ccall, \callpp, \cdr)$ on behalf of $\adv$ to $\fprivtb$. 
    \item Receive $(\ccall, \ctime)$ from $\fprivtb$ and store $(\ctime, (\adv, \callpp, \cdr))$ in $\DDD$. 
\end{enumerate}

\myparagraph{\prototrace} 
We consider two cases: 1) an honest trace request 2) a malicious trace request 

\begin{enumerate} 
\item Honest trace request: Upon receiving $(\ctrace,\ctime)$ from $\fprivtb$
\begin{enumerate}
    \item If $(\ctime, (\adv, \callpp, \cdr)) \in \DDD$
    \begin{enumerate}
        \item Sample random $r \gets \ZZ_q$ and compute $a = H_1(\callpp)^r$ and send $a$ to $\adv$
        \item Receive $b$ from $\adv$. 
        \item Check that $e(\pk_\oprf, H_1(\callpp)) = e(g, b^{1/r})$.      
        \item Output $\clabel^* = H_2(\pk_\oprf, \callpp, b^{1/r})$ 
        \item Request $\adv$ for a signature on $\clabel^*$. 
        \item If no signature received send $\emptyset$ to $\fprivtb$. And upon receiving $(\allowtrace, P_i)$ from $\fprivtb$, return $(\allowtrace, \bot)$
        \item Else Decrypt $\ct_1, \ct_2$ corresponding to $\clabel^*$ if it exists and send ${\ctime, \adv, \callpp, \cdr}$ to $\fprivtb$. And upon receiving $(\allowtrace, P_i)$ from $\fprivtb$, return $(\allowtrace, \ok)$
        \item If no ciphertexts exist corresponding to this $\clabel^*$ abort with failure $\veroprffail$.
        
    \end{enumerate}
    \item If $(\ctime, (\adv, \callpp, \cdr)) \notin \DDD$, 
    \begin{enumerate}
        \item simulate the computation of a label as in honest contribution
        \item Sample a random string $\clabel^*$ . 
        \item Request $\adv$ for a signature on $\clabel^*$. If a signature is received, send $(\allowtrace, \ok)$ to $\fprivtb$, else send $(\allowtrace, \bot)$ upon receiving $(\allowtrace, P_i)$ from $\fprivtb$. 
    \end{enumerate}
\end{enumerate}

\item Malicious trace request: 
\begin{enumerate}
    \item Upon receiving $(\clabel^*, \sigma_i, \sigma_R)$ from $\adv$ (on behalf of $P_j^*$): 
    \begin{enumerate}
    \item If $\sigma_i$ corresponds to that of an honest party, abort with $\sigfail$. 
        \item If $\clabel^*, (\cdot)$ exists in the list of ciphertexts, send $(\clabel^*, (\ct_1, \ct_2, \sigma))$ to $\adv$. Else:
        \item Check $(x, \clabel)$ exists in $Q_2$. If not, abort with $\rofail_2$. 
        \item Else parse $x$ as $(\pk_\oprf, \callpp, b^*)$ 
        \item Check that $(\callpp, y)$ exists in $Q_1$. If not, abort with $\rofail_1$. 
        \item Check that $e(\pk_\oprf, y) = e(g, b^*)$. If not, abort with $\prffail$
    \end{enumerate}
    \item Send $(\ctrace, \callpp, P_j^*)$ on behalf of $P_j^*$ to $\fprivtb$ 
    \item Receive $(\{\ctime\})$ from $\fprivtb$. If any of the $\ctime$ correspond to that of a malicious contribution, send $\{\ctime, P_j, \callpp, \cdr\}$ to $\fprivtb$. 
    \item Upon receiving $\allowtrace$ from $\fprivtb$,  send \\$(\allowbreak\allowtrace, \ok)$ to $\fprivtb$. 
    \item Receive $\CCC_\trace$ from $\fprivtb$. 
    \item Encrypt each of the $\cdr$ using the corresponding $\clabel$ and $\vk_T$ in the following way:
    \begin{enumerate}
        \item Sample a random key $\cdrkey \gets \{0,1\}^\lambda$ 
        \item Compute $\ct_1 = \WE.\enc((\clabel, \vk_T), \cdrkey)$
        \item  Sample a random  $\ct_2$. Set $H_3(\callpp\|\cdrkey) = \ct_2\oplus \cdr$
    \end{enumerate}
    \item Using the group signature scheme, compute $\sigma$ on behalf of $P_i$, where $P_i$ is the honest sender of the record. 
    \item Compute $\sigma_{RS} = \sign((\clabel^*, \ct_1^*, \ct_2^*, \sigma^*))$
    \item Send the $(\clabel^*, \ct_1^*, \ct_2^*, \sigma^*), \sigma_{RS}$ that correspond to $\clabel^*$ to the adversary. 
\end{enumerate}

\end{enumerate}

\myparagraph{\protoopen} Upon receiving $(\copen, \callpp, \CCC_\trace)$ request from $\fprivtb$  send $\gopen$ request to $\adv$ and output whatever $\adv$ returns. 
\begin{enumerate}
    \item If the adversary opens a signature submitted by a malicious party as an honest party's identity abort with $\framefail$. 
    \item If the adversary opens a record $(\clabel^*, \ct_1^*, \ct_2^*, \sigma^*), \sigma_{RS}$ where $\sigma_{RS}$ was not computed by the simulator abort with $\sigfail_2$. 
\end{enumerate}

\myparagraph{Proof By Hybrids} Now to prove that the simulated world and the real world are indistinguishable we proceed via a sequence of hybrids, starting from the real world until we reach the ideal world. We show that each of these hybrids are indistinguishable and therefore the real world and the simulated world are indistinguishable. 

\begin{itemize}
    \item[$\hyb_0$] This is the real world protocol 
    \item[$\hyb_1$] This hybrid is identical to the previous hybrid except that the simulator may abort with $\groupsigfail$. By the non-frameability property of the group signature scheme, the simulator aborts with negligible probability and therefore this hybrid is indistinguishable from the previous one. 
    \item[$\hyb_2$] This hybrid is identical to the previous hybrid except that the simulator may abort with $\rofail_1$. Since we use a random oracle and require the adversary to use the RO, the probability of this event occurring is negligible, and therefore this hybrid is indistinguishable from the previous one. 
    \item[$\hyb_3$] This hybrid is identical to the previous hybrid except that the simulator may abort with $\rofail_2$.Since we use a random oracle and require the adversary to use the RO, the probability of this event occurring is negligible, and therefore this hybrid is indistinguishable from the previous one. 
    \item[$\hyb_4$] This hybrid is identical to the previous hybrid except that the simulator may abort with $\rofail_3$. Since we use a random oracle and require the adversary to use the RO, the probability of this event occurring is negligible, and therefore this hybrid is indistinguishable from the previous one. 
    \item[$\hyb_5$] This hybrid is identical to the previous hybrid except that the simulator may abort with $\sigfail$. Since we use unforgeable signatures, this event occurs with negligible probability and therefore the two hybrids are indistinguishable. 
    \item[$\hyb_6$] This hybrid is identical to the previous hybrid except that the simulator may abort with $\veroprffail$. Since we use a verifiable OPRF scheme this occurs with negligible probability. 
    \item[$\hyb_7$] This hybrid is identical to the previous hybrid except that the simulator simulates the OPRF calls using 0 as input instead of the $\callpp$. By the obliviousness property of the underlying OPRF scheme, these two hybrids are indistinguishable. 
    \item[$\hyb_8$] This hybrid is identical to the previous hybrid except that the simulator may abort with $\sigfail_2$. Since we use unforgeable signatures, this event occurs with negligible probability and therefore the two hybrids are indistinguishable. 
\end{itemize}

Since this hybrid is identical to the simulated world, we have shown that the real world and ideal world are indistinguishable, and that concludes the proof of security for the case when only a subset of carriers are corrupt. 

\subsection{Case 3: \RS and a subset of the providers are corrupt, and \TA is honest} 

\myparagraph{\protosetup}
Simulate the  Traceback Authority: 
\begin{enumerate}
    \item Generate BLS signature keys $(\sk_T, \vk_T) \gets \sign.\kgen(1^\lambda)$  and $(\sk_R, \vk_R) \gets \sign.\kgen(1^\lambda) $
    \item Generate OPRF key $k \gets \ZZ_q$ and announce the corresponding public key $\pk_\oprf = g^k$
    \item Run the $\gkgen$ algorithm of the group signature scheme and announce $(\gpk, \info_0)$  which are the group manager public key and  the initial group information. 
\end{enumerate}

Simulate honest providers: Upon receiving $(\cregister, P_i)$ from $\fprivtb$: Generate signing keys $(\sk_i, \vk_i) \gets \sign.\kgen(1^\lambda)$ and announce $\vk_i$. 

Malicious provider $P_j$: Simulate the interactive joining protocol $\gjoin$ with $P_j$, and send $(\cregister, P_i)$ to $\fprivtb$. 

\myparagraph{\protocontribute}

Simulating honest contributions: Upon receiving $(\ccall, \ctime)$ from $\fprivtb$,
\begin{enumerate}
    \item Sample a random $\clabel$
    \item Send a random $\cdrkey$ and compute $\ct_1 = \WE.\enc((\vk_T, \clabel), \cdrkey)$
    \item Sample a random string $\ct_2$ 
    \item Compute a group signature $\sigma$ on behalf of some honest party $P_i$
    \item Send $(\ct_1, \ct_2, \clabel, \sigma)$ to $\RS$ and send $(\ccall, \ok)$ to $\fprivtb$. Store $(\ctime, \clabel, \ct_1, \ct_2)$. 
\end{enumerate}

Simulating random oracle invocations: 
\begin{enumerate}
    \item Upon receiving input $x$ for random oracle $H_1$, check if $(x, y)$ exists in $\QQQ_1$. If yes return $y$, else sample a random $y$, store $(x,y) \in \QQQ_1$ and return $y$. 
    \item Upon receiving input $x$ for random oracle $H_2$, check if $(x, y)$ exists in $\QQQ_2$. If yes return $y$, else sample a random $y$, store $(x,y) \in \QQQ_2$ and return $y$. 
     \item Upon receiving input $x$ for random oracle $H_3$, check if $(x, y)$ exists in $\QQQ_3$. If yes return $y$, else sample a random $y$, store $(x,y) \in \QQQ_3$ and return $y$. 
\end{enumerate}

Simulating malicious contributions: Note that since the $\RS$ is corrupt, the simulator only needs to simulate the label generation. 
\begin{enumerate}
    \item Upon receiving $a$ on behalf of $\TA$ from the adversary, compute $b = a^k$ and send it back to the $\adv$. 
\end{enumerate}

\myparagraph{\prototrace} 
We consider two cases: 1) an honest trace request 2) a malicious trace request 

\begin{enumerate} 
\item Honest trace request: 
\begin{enumerate}
    \item Upon receiving $(\ctrace,\ctime, \callpp)$ from $\fprivtb$,
    \begin{enumerate}
        \item Retrieve $\clabel$  that corresponds to $\ctime$ if it exists. 
        \item Send $(\clabel, \sigma_R)$ to $\RS$ and receive $(\clabel, \ct_1, \ct_2, \sigma)$. If no such record was received, send $(\malupdate, (\delete, \ctime)$ to $\fprivtb)$
        \item If no $\clabel$ exists, compute $\clabel = H_2(\pk_\oprf,\allowbreak \callpp, H_1(\callpp)^k)$ and send $\clabel$ to $\adv$. If any $\ct_1, \ct_2, \clabel,\sigma$ received, first check if $\sigma$ corresponds to that of an honest party, if this is the case abort the simulation with $\groupsigfail$ else decrypt the ciphertexts to retrieve the $\cdr$ and send $(\malupdate, (\add,\allowbreak (P_j^*, \callpp, \cdr)))$ to $\fprivtb$. 
        Upon receiving $\ctime$, send ${\ctime, P_j^*, \callpp,\cdr}$ to $\fprivtb$. 
\end{enumerate}
    \item Upon receiving $(\allowtrace, P_i)$ from $\fprivtb$, send $(\allowtrace, P_i, \ok)$ back to $\fprivtb$. 
\end{enumerate}

\item Malicious trace request:
\begin{enumerate}
\item Upon receiving a query for $H_3$ on $\callpp\|\cdrkey$, send $(\ctrace, \callpp, P_i)$ to $\fprivtb$ and receive back $\CCC_\trace = \{\ctime, \cdr\}$. 
    \item  Retrieve $(\ct_1, \ct_2, \clabel, \sigma)$ that correspond to $\ctime$ and check that this is the same $\cdrkey$ that was encrypted in $\ct_1$. If yes, compute $z = \ct_2 \oplus \cdr$ and send $z$ in response and store $((\callpp\|\cdrkey), z)$ in $\QQQ_3$.  If not, sample a random $z \gets \{0,1\}^\lambda$ and send $z$ in response. 
\end{enumerate}

\end{enumerate}

\myparagraph{\protoopen} Upon receiving $\gopen$ request from $\adv$ for a particular record, send $\copen, \callpp, \CCC_\trace$ to $\fprivtb$ and output whatever $\fprivtb$ returns.

\myparagraph{Proof By Hybrids} Now to prove that the simulated world and the real world are indistinguishable we proceed via a sequence of hybrids, starting from the real world until we reach the ideal world. We show that each of these hybrids are indistinguishable and therefore the real world and the simulated world are indistinguishable. 

\begin{itemize}
    \item[$\hyb_0$] This is the real world protocol 
    \item[$\hyb_1$] This hybrid is identical to the previous hybrid except that the simulator may abort with $\groupsigfail$. By the non-frameability property of the group signature scheme, the simulator aborts with negligible probability and therefore this hybrid is indistinguishable from the previous one. 
    \item[$\hyb_2$] This hybrid is identical to the previous hybrid except that the simulator may abort with $\rofail_1$. Since we use a random oracle and require the adversary to use the RO, the probability of this event occurring is negligible, and therefore this hybrid is indistinguishable from the previous one. 
    \item[$\hyb_3$] This hybrid is identical to the previous hybrid except that the simulator may abort with $\rofail_2$.Since we use a random oracle and require the adversary to use the RO, the probability of this event occurring is negligible, and therefore this hybrid is indistinguishable from the previous one. 
    \item[$\hyb_4$] This hybrid is identical to the previous hybrid except that the simulator may abort with $\rofail_3$. Since we use a random oracle and require the adversary to use the RO, the probability of this event occurring is negligible, and therefore this hybrid is indistinguishable from the previous one. 
    \item[$\hyb_5$] This hybrid is identical to the previous hybrid except that the simulator may abort with $\sigfail$. Since we use unforgeable signatures, this event occurs with negligible probability and therefore the two hybrids are indistinguishable. 
    \item[$\hyb_6$] This hybrid is identical to the previous hybrid except that the simulator simulates the honest contributions by sampling a random $\clabel$. By the pseudorandomness property of the underlying OPRF scheme, these two hybrids are indistinguishable. 
    \item[$\hyb_7$] This hybrid is identical to the previous hybrid except that the ciphertext $\ct_2$ is a randomly sampled string. By the perfect security of the stream cipher (OTP), these two hybrids are indistinguishable. 
    \item[$\hyb_8$] This hybrid is identical to the previous hybrid except that the group signature corresponds to that of a random honest party. By the anonymity guarantees of the underlying group signature scheme, these two hybrids are indistinguishable. 
\end{itemize}

Since this hybrid is identical to the simulated world, we have shown that the real world and ideal world are indistinguishable, and that concludes the proof of security for the case when only a subset of carriers are corrupt.      \begin{sloppypar}

\section{Artifact Appendix}

\subsection{Abstract}
We developed a prototype of the \sysname{} system and conducted a thorough performance evaluation. The \sysname{} artifact is composed of four key components: Group Membership Management, Label Generation, Trace Authorization, and Record Storage. Each of these components was containerized using Docker, and we orchestrated them together with Docker Compose. Additionally, we integrated auxiliary services, including a web-based GUI, to facilitate interaction with the database.

We generated a dataset of Call Detail Records and evaluated \sysname's performance. Our experimental results indicate that \sysname{} incurs minimal computational and bandwidth overhead per call, with these costs scaling linearly with the increase in call volume.

\subsection{Description \& Requirements}
The \sysname{} prototype comprises four integral components:

\myparagraph{Membership Management}
The Group Manager (GM) oversees membership management. This component enables the GM to issue new group membership keys or revoke existing ones, and it also facilitates the tracing of traitors. Within our implementation, the \TA{} assumes the role of the GM.

\myparagraph{Label Generation}
Label generation is controlled by the Label Manager (LM). The LM collaborates with providers to evaluate pseudorandom functions using the Oblivious Pseudorandom Function protocol. In our system, the \TA{} also fulfills the role of the LM.

\myparagraph{Trace Authorization}
This component is responsible for generating authorization signatures required to decrypt ciphertexts. The \TA{} acts as the Trace Authorizer in our implementation.

\myparagraph{Record Storage}
The Record Storage component stores ciphertexts and provides the results of match trace queries.

\vspace{5pt}\noindent All the above components are implemented in Python. However, for performance optimization, we implemented the witness encryption in C++ and created Python bindings to integrate the library into our prototype.

\subsubsection{Security, privacy, and ethical concerns}
None

\subsubsection{How to access}
We have archived the witness encryption and the \sysname{} prototype into a zip file, which is publicly accessible on Zenodo at the following link: \href{https://zenodo.org/doi/10.5281/zenodo.12733869}{https://zenodo.org/doi/10.5281/zenodo.12733869}. Additionally, we maintain an active version of the artifact in our GitHub repositories. The source code for Witness Encryption can be found at \href{https://github.com/wspr-ncsu/BLS-Witness-Encryption}{https://github.com/wspr-ncsu/BLS-Witness-Encryption}, while the \sysname{} prototype is available at \href{https://github.com/wspr-ncsu/jaeger}{https://github.com/wspr-ncsu/jaeger}.

\subsubsection{Hardware dependencies}
Running \sysname{} does not necessitate any specific hardware requirements. However, to achieve results comparable to those presented in the paper, our experiments were conducted on a Linux virtual machine equipped with 32 vCPUs and 64 GB of memory. The underlying host was a Super Micro Server featuring an Intel Xeon Gold 6130 processor, ECC DDR RAM, and 12 Gbps SAS drives.

\subsubsection{Software dependencies}
For ease of setup, we recommend configuring the project using Docker. If you prefer not to use Docker, our repositories provide detailed instructions on how to set up the project without it. For the remainder of this artifact appendix, we will focus exclusively on the setup and execution of experiments using Docker.

\subsubsection{Benchmarks}
None

\subsection{Set Up}
\subsubsection{Installation} 
You need to install the appropriate version of Docker based on your operating system. If your Docker installation does not include the Docker Compose plugin, be sure to install Docker Compose separately. Additionally, download the \sysname{} prototype source code from either the GitHub repository or Zenodo.

We have published the \sysname{} Docker image on Docker Hub as \texttt{kofidahmed/jager}. If this image is not available, navigate to the root directory and build the \sysname{} Docker image using the following command:
\begin{verbatim}
docker build -t kofidahmed/jager .
\end{verbatim}

\noindent Note that the \texttt{-t} option does not necessarily need to be \texttt{kofidahmed/jager}. We use \texttt{kofidahmed/jager} to align with the image on dockerhub.

\subsubsection{Basic test}\label{ae:basic-test}
Generate secret keys for label generation, group master and public keys for group management, as well as private and public keys for BLS signatures and witness encryption by running the following command:

\begin{verbatim}
docker run \ 
  -v $(pwd):/app \
  --rm kofidahmed/jager \
  python keygen.py -a
\end{verbatim}

\noindent The \texttt{-a} option instructs the script to generate all necessary keys. If you only need to generate keys for specific components, use the following options: \texttt{-lm} for label generation, \texttt{-gm} for group management, and \texttt{-ta} for trace authorization/witness encryption. After running the command, verify that the \texttt{.env} and \texttt{membership-keys.json} files have been created and that the variables within are populated with the appropriate keys.

\subsection{Evaluation Workflow}

\subsubsection{Major claims}
\begin{compactitem}
\item[(C1):] 
The average runtime for the following operations are as follows: Label generation takes 0.073 ms, the contribution protocol takes 4.143 ms, trace authorization takes 0.419 ms, decryption takes 0.847 ms, opening a signature takes 0.147 ms, and verifying a group signature takes 2.310 ms. Experiment (E1) substantiates these performance metrics.

\end{compactitem}

\subsubsection{Experiments}
\begin{compactitem}
\item[(E1):] Benchmark \sysname{} operations in Table \ref{table:rec-perf}.
  \begin{asparadesc}
  \item[Preparation:] Run the Docker Compose command to start the services, and then log in to the Docker container by executing: 
\begin{verbatim}
  docker compose up -d
  docker exec -it jager-exp bash
\end{verbatim}

\item[Execution:] To benchmark the operations, run the following command: 
\begin{verbatim}
  python benchmarks.py -a
\end{verbatim}
  
This will display the results on the console and create a \texttt{results} folder inside the project root.  

  \item[Results:] 
      The file \texttt{results/bench.csv} contains a summary of the benchmarks, while \texttt{results/index-timings.csv} records the individual runs. We used \texttt{results/index-timings.csv} to determine the mean, min, max, and standard deviations. To aggregate the benchmark results from \texttt{results/index-timings.csv}, as shown in Table \ref{table:rec-perf} (in the paper), run \texttt{python aggregate-benchmark.py}. This script generates a CSV file \texttt{results/bench-summary.csv} with the aggregated results.
\end{asparadesc}

\item[(E2):] Determine Bandwidth, Storage Growth, and Query Performance as illustrated in Fig. \ref{fig:query-perfs}.
    \begin{asparadesc}
        \item[Preparation:] Data generation 
            \begin{enumerate}
                \item Run Docker Compose to start the services, and log in to the Docker container as demonstrated in E1.
                \item Generate telephone and social network data by running the command: \texttt{python datagen.py -n 100 -s 10000 -c -y}. The \texttt{-n} option specifies the number of carriers, \texttt{-s} specifies the number of subscribers, \texttt{-c} determines whether CDRs should be generated, and \texttt{-y} skips all prompts. Note that in the paper, \texttt{-n} is set to 7000 and \texttt{-s} is set to 300M.
                \item View the Generated Dataset: We will connect to the ClickHouse database using a web browser. Ensure that your browser has network access to both ports 5521 and 8123. We have added a UI service that allows you to connect to the database. Visit \texttt{http://localhost:5521} in your browser.
                \begin{itemize}
                    \item Enter \texttt{http://localhost:8123} as the ClickHouse URL, \texttt{default} as the Username, and \texttt{secret} as the Password, then click the \texttt{Submit} button. Once successful, click the \texttt{Go back to the home page} link.
		          \item On the home page, select \texttt{Jager} in the database field, which will load the tables. You can then click on any table to view its Details, Schema, or preview rows. 
		          \item If you prefer to run your own SQL queries, click on the new file icon/button with the orange background. Type \texttt{select * from jager.raw\_cdrs limit 10;} in the query field and click the \texttt{Run Query} button. 
                \end{itemize}
            \end{enumerate}
        \item[Execution:] To run the contributions protocol, execute the command: \texttt{python run-contribution.py -b 3 -r 100}. The \texttt{-b} option specifies the number of batches, which is the number of times you wish to run the contribution experiment (it defaults to \texttt{1}). After each batch, database stats are measured and stored in the \texttt{results} folder. The \texttt{-r} option is required and specifies the number of records per batch.
        \item[Results:] The results are saved in \texttt{results/db\_stats.csv} and \texttt{results/queries.csv}. To generate the fetch and insert query performance graph shown in Fig. \ref{fig:query-perfs} from the results in \texttt{results/queries.csv}, run \texttt{python plot-db-stats.py}. This creates a PNG file at \texttt{results/query\_performance.png}.
    \end{asparadesc}

\item[(E3):] Run a Traceback (Optional)
    \begin{asparadesc}
        \item[Preparation:] Start the Docker services and log in to the Docker container as described in E1. Visit \texttt{http://localhost:8123} in your browser, and execute the SQL command \texttt{select src, dst, ts from jager.raw\_cdrs where status = 1;}. Refer to the E2 preparation section for instructions on executing the SQL query. The results from this query are the calls whose ciphertexts have been submitted to the RS.
        
        \item[Execution:] Run the following command: \texttt{python run-trace.py -s src -d dst -t ts}. Replace \texttt{src}, \texttt{dst}, and \texttt{ts} with the values from any row in the query results obtained in the preparation step above.
        
        \item[Results:] This command will:
            \begin{itemize}
    		\item Generate call labels within the $[\ts - \tmax, \ts + \tmax]$ range
    		\item Request authorization signatures from the TA
    		\item Retrieve ciphertexts from the RS
    		\item Decrypt and analyze the records to determine the origin and call path.
                \item Display the results to your console.
            \end{itemize}
    \end{asparadesc}
\end{compactitem}
\bigskip

\subsection{Notes on Reusability}
Our implementation includes Docker services defined in the \texttt{compose.yml} file. Once the compose-up command is running, the following services are exposed via \texttt{http://localhost:\{PORT\}}:
\begin{itemize}
    \item The Group Management server runs on \texttt{9990}. The implementation is defined in \texttt{app-gm.py}.
    \item The Label Generation server runs on \texttt{9991}. The implementation is defined in \texttt{app-lm.py}.
    \item The Trace Authorization server runs on \texttt{9992}. The implementation is defined in \texttt{app-ta.py}.
    \item The Record Store server runs on \texttt{9993}. The implementation is defined in \texttt{app-rs.py}.
\end{itemize}

\noindent For more information on the commands available for customizing our implementation, please refer to our GitHub repository.

\subsection{Version}
Based on the LaTeX template for Artifact Evaluation V20220926.

     \end{sloppypar}
\fi

\end{document}